\documentclass[showpacs,superscriptaddress,twocolumn,prb]{revtex4}
\usepackage{amsmath,graphicx,float}

\begin{document}

\title {High Frequency Quantum Admittance and Noise Measurement with an On-chip Resonant Circuit }
\author{J. Basset}
\affiliation{Laboratoire de Physique des Solides, Universit\'e Paris-Sud, CNRS, UMR 8502, F-91405 Orsay Cedex, France.}
\author{H. Bouchiat}
\affiliation{Laboratoire de Physique des Solides, Universit\'e Paris-Sud, CNRS, UMR 8502, F-91405 Orsay Cedex, France.}
\author{R. Deblock}
\affiliation{Laboratoire de Physique des Solides, Universit\'e Paris-Sud, CNRS, UMR 8502, F-91405 Orsay Cedex, France.}

\begin{abstract}
By coupling a quantum detector, a superconductor-insulator-superconductor junction, to a Josephson junction \textit{via} a resonant circuit we probe the high frequency properties, namely the ac complex admittance and the current fluctuations of the Josephson junction at the resonant frequencies. 
The admittance components show frequency dependent singularities related to the superconducting density of states while the noise exhibits a strong frequency dependence, consistent with theoretical predictions. The circuit also allows to probe separately the emission and absorption noise in the quantum regime of the superconducting resonant circuit at equilibrium. At low temperature the resonant circuit exhibits only absorption noise related to zero point fluctuations, whereas at higher temperature emission noise is also present.
\end{abstract}

\maketitle 

\section{Introduction}
\label{INTRO}

Transport and current fluctuations measurements are powerful tools to study mesoscopic conductors. Whereas those quantities have been extensively studied at low frequency in various systems \cite{imry97,blanter00} investigation of the conductance and noise at high frequency is much more recent. Measuring those quantities allows to probe the dynamics and the fluctuations of a mesoscopic system in a frequency range of the order of or higher than the inverse internal timescales of the system or the applied voltage and temperature characteristic energy scales. 
Indeed ac complex conductance measurements on diffusive coherent wires \cite{pieper92} and rings \cite{pieper94} bears signature of the dephasing time. For a superconducting junction it reveals a reactive quasiparticle singularity related to the superconducting density of state \cite{hu90,worsham91}. On the other hand, the current fluctuations in the quantum regime, where the frequency is of the order of or higher than temperature or voltage bias characteristic energy scale, acquire a frequency dependence with signatures of the relevant energy scales $k_B T$ and $eV$ (with $T$ the temperature and $V$ the bias voltage on the device) and has been found to increase linearly with frequency above $k_BT/h$ \cite{koch81,yurke88}.  Similarly  the  excess noise, \textit{i.e.} the difference between the noise at a given bias and the noise at equilibrium, measured in the limit $eV \gg k_BT$  has been found to decrease linearly with frequency and go to zero at frequency $eV/h$ both in diffusive wires \cite{schoelkopf97} and GaAs ballistic quantum point contacts \cite{eva07}. 

The ac conductance is the current response of the system to an ac voltage excitation whereas current noise corresponds to the current fluctuations associated to a dc voltage applied on the system. Those two quantities are related by the generalized\footnote{Originally the fluctuation-dissipation theorem describes the properties of dissipative systems at equilibrium \cite{kubo66}. This is only very recently than this theorem was generalized to out-of-equilibrium situations \cite{safi08,safi09}.} fluctuation-dissipation theorem \cite{kubo66,safi08,safi09} in the quantum regime ($h\nu>>k_BT$). In this regime noise can be described in terms of exchange of photons of energy $h \nu$ between the source and the noise detector. Depending on whether photons are emitted or absorbed by the source one measures emission noise (corresponding to negative frequencies) and absorption noise (corresponding to positive frequencies) \cite{clerk10}. The generalized fluctuation-dissipation theorem states that the difference between the absorption noise and the emission noise at a given frequency is proportional to the dissipative part of the conductance at this frequency. Measuring simultaneously on the same system the current fluctuations, both in emission and in absorption, and the complex ac conductance is thus of particular interest. 

The difference between emission and absorption processes is well known in the field of quantum optics but difficult to observe in electronic devices since most classical amplifiers exchange energy with the device measured and allow only the detection of a combination of emission and absorption noise \cite{lesovik97}. On the other hand a quantum detector \cite{clerk10,aguado00}, such as a superconductor-insulator-superconductor (SIS) tunnel junction \cite{deblock03,billangeon06,basset10}, allows to measure the non-symmetrized noise, \textit{i.e.} distinguish between emission and absorption.

In this work we demonstrate that by incorporating a SIS quantum detector and the tested device, a Josephson junction, in an on-chip superconducting resonant circuit it is possible to extract the complex admittance at finite frequency of the source junction from the measurement of the $I(V)$ characteristic of the detector. The real and imaginary part of the complex admittance are measured at the resonant frequencies of the resonator and exhibit frequency dependent singularities related to the superconducting density of states. With the same setup we access the excess emission noise spectral density of quasiparticles tunneling through the Josephson junction. It exhibits a strong frequency dependence as predicted by theory. We also show that using the same circuit it is possible to detect the emission and absorption noise of the resonant circuit at \textit{equilibrium}. In particular, at the frequencies probed in the experiment ($28.4$ and $80.2$ GHz, the resonant frequencies of the resonator), at low temperature the resonator does not emit noise wheras it shows absorption noise related to its zero point fluctuations. Emission and absorption can be quantitatively related to the real part of the high frequency impedance of the resonant circuit. This latter quantity is calibrated with the same setup using ac Josephson effect. 

This detection scheme is ideally suited for measuring high frequency properties of device which resistance are much higher than the impedance of the resonant circuit. Thus the device resistance will be typically in the range of the resistance quantum. This condition ensures that the properties of the resonator are not too much affected by the presence of the system under test and the detector. Due to the need to incorporate both the tested device and the SIS detector in the resonant circuit, this technique is best suited for nanoscale devices.    

The paper is organized as follows. Section \ref{SFAMT} is devoted to the description of the experimental setup and its modelling. It allows to determine the parameters affecting the resonant frequencies and quality factors of the resonant circuit, the key ingredients of the detection schemes. Section \ref{QA} is devoted to the measurement of the complex quantum admittance of a Josephson junction at finite frequency. Section \ref{SISNoiseDetector} details the theoretical principle of noise detection with an SIS junction at high frequency while section \ref{NoiseJoseph} records the extraction of the emission noise power of quasiparticles tunneling through the Josephson junction. Finally, section \ref{QNMWOCRC2} shows a separate measurement of the emission and absorption noise in the quantum regime of the superconducting resonant circuit at equilibrium verifying the fluctuation-dissipation theorem.

\section{Sample Fabrication and Modelling}
\label{SFAMT}

We aim at coupling a noise source to a quantum noise detector at high frequency (GHz range) with the capability to bias them independently at low frequency. The first approach chosen in previous works \cite{deblock03,billangeon06} consisted in a capacitive coupling. We choose in this work to use a resonant coupling circuit to improve the source/detector coupling and get more quantitative results.
In this section we present the sample and model it. It allows to determine the effect of different parameters on the coupling between the noise source and the detector.

\subsection{Sample Fabrication}

\begin{figure}[htbp]
  \begin{center}
		\includegraphics[width=8cm]{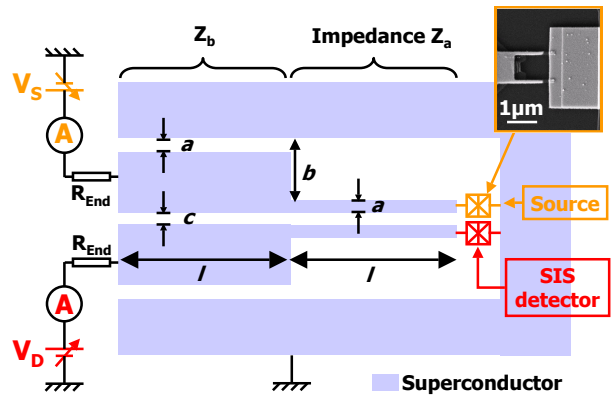}
	\end{center}
  \caption{Sketch of the sample, which is constituted of two coupled transmission lines. Each line is connected to the ground via a DC SQUID constituted by two Josephson junctions in parallel. One of the SQUID, with its critical current possibly minimized by a magnetic flux, is considered as a SIS detector (or detector junction) whereas the other SQUID is called the source junction. The dimensions of the sample are $a=5\mu m$, $b=100\mu m$, $c=5\mu m$ and $l=1mm$. The two transmission lines are terminated by two on-chip Pt resistors in parallel resulting in $R_{End}=413 \Omega$. The junctions are made by shadow angle evaporation. The DC SQUIDs have different loop size in order to tune separately their Josephson currents.}
  \label{fig1}
\end{figure}

The device probed in this experiment consists of two coupled coplanar transmission lines. Each transmission line is connected to the ground plane via a DC squid, constituted by two small Josephson junctions in parallel, and consists of two sections of same length $l$ but with different width, thus different characteristic impedance $Z_a \approx 110 \Omega$ and $Z_b \approx 25 \Omega$ (Fig. \ref{fig1}). Due to the impedance mismatch the transmission line acts as a quarter wavelength resonator, with resonances at frequency $\nu_n=nv/4l=n\nu_1$, where $v$ is the propagation velocity and $n$ an odd integer \cite{holst94}. The two transmission lines are fabricated close to one another to provide a good coupling at the resonant frequencies and are terminated by two on-chip Pt resistors ($150 \times 2 \times 0.03 \mu m^3$) in parallel. The value of each resistance is $2R_{End}=826\Omega$. The size of the Josephson junctions of the DC SQUIDs is $240\times150$ nm$^2$. The two DC SQUIDs have different size loop (1 and 6 $\mu$m$^2$) in order to tune separately their critical currents with a magnetic flux. The small size of the junctions and the small geometrical inductance of the loop insure that the SQUID can be considered as a single junction with a tunable Josephson energy \cite{lefevre92}. The junctions and the resonator are fabricated in aluminum ($Al(30nm)/AlO_x/Al(50nm)$, superconducting gap $\Delta =260 \mu$eV) on a high resistivity oxidized silicon wafer. The system is thermally anchored to the cold finger of a dilution refrigerator of base temperature 20 mK and measured through filtered lines with a standard low frequency lock-in amplifier technique. In the following we call one junction the detector junction and  the other the source. The detector has a normal state resistance $R_N=18.7k\Omega$ and thus a maximum critical current of $I_C=\pi \Delta/(2eR_N)=21.8$nA \cite{tinkham96,ambegaokar63}. The source normal state resistance is $19.5k\Omega$.

\subsection{Coupled transmission lines description}
\label{MODELISATION}

We now model the two coupled transmission lines in order to introduce the impedance matrix describing the coupling circuit between the source and the detector.

\subsubsection{Model}
  
\begin{figure}[htbp]
  \begin{center}
		\includegraphics[width=8cm]{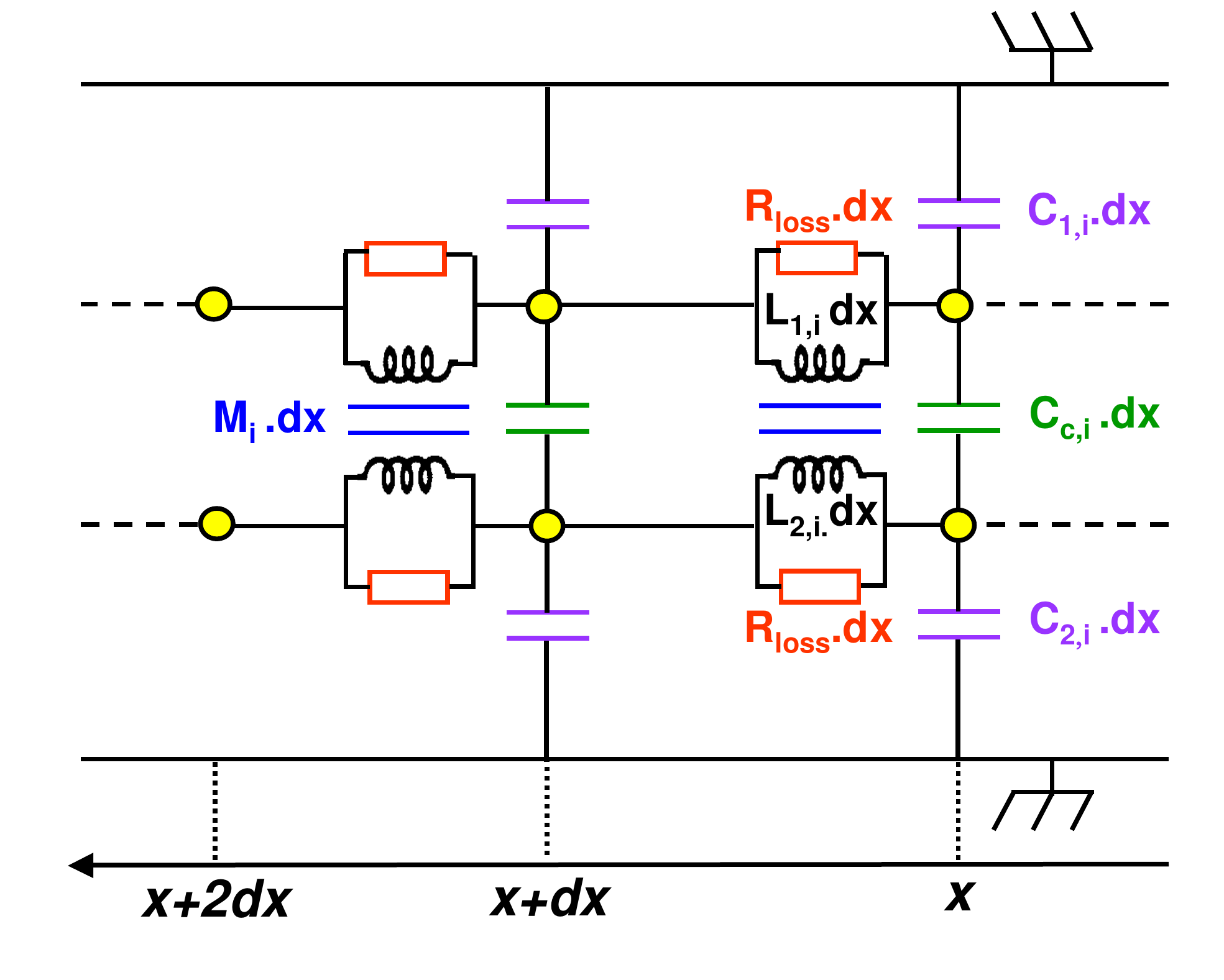}
	\end{center}
  \caption{Model circuit used to extract the eigen frequencies of the two coupled transmission line resonators (designed by numbers $1$ and $2$) and estimate the quality factors. It consists in two distributed LC circuits (distributed inductance $L_{1,i}$, $L_{2,i}$ and capacitance $C_{1,i}$, $C_{2,i}$) capacitively and inductively coupled (distributed mutual inductance $M_{i}$ and coupling capacitance $C_{c,i}$). The index i denotes which section of the transmission line is considered \textit{i.e} section a or b (see Fig.\ref{fig1}). They have different values of distributed inductance and capacitance leading to an impedance mismatch in $x=l$. The values are fixed by the geometry and the choice of materials.}
  \label{Circuit}
  \end{figure}

The sample is modeled as two distributed LC transmission lines capacitively and inductively coupled via the distributed coupling capacitance $C_C$ and the distributed mutual inductance $M$ as shown in Fig.\ref{Circuit}. Both transmission lines have an impedance mismatch at $x=l$. We call $V_1(x)$ and $V_2(x)$ the voltages developing along the transmission line 1 (top) and 2 (bottom) and $i_1(x)$ and $i_2(x)$ the currents flowing into these same lines. The equations of propagation along the line 1 read at frequency $\nu=\omega/2\pi$:
\begin{equation}
		\left\{
		\begin{array}{rl}
			\frac{d^2V_1}{dx^2}=\Gamma_{+}^2V_1+\Gamma_{-}^2V_2\\
			\frac{dV_1}{dx}=-j\omega L' i_1-j\omega M i_2\\
		\end{array}
		\right.
\end{equation}

with $L'=LR_{Loss}/(R_{Loss}+j\omega L)$, $\Gamma_{+}^2=j\omega L'.j\omega (C+C_C)+j\omega M.j\omega C_C$, $\Gamma_{-}^2=j\omega M.j\omega (C-C_C)-j\omega L'.j\omega C_C$ and $j^2=-1$. $R_{Loss}$ is the value of the distributed resistance of the transmission line, $L$ the distributed inductance and $C$ the distributed capacitance. The equation for line 2 can be deduced by exchanging 1 and 2 in the previous expressions.

\subsubsection{Eigen modes of the resonator}

We now introduce the eigen modes of the resonator which are \textit{even} and \textit{odd} modes defined respectively as $V_E=(V_1+V_2)/2$,  $V_O=(V_1-V_2)/2$, $i_E=(i_1+i_2)/2$ and  $i_O=(i_1-i_2)/2$ . One can then read $V_1=V_E+V_O$ and $V_2=V_E-V_O$. The previous equations simplify to :

\begin{equation}
		\left\{
		\begin{array}{rl}
			\frac{d^2V_E}{dx^2}=\Gamma_E^2V_E; \frac{d^2V_O}{dx^2}=\Gamma_0^2V_O \\
			\\
			\frac{dV_E}{dx}=-j\omega (L'+M))i_E; \frac{dV_O}{dx}=-j\omega (L'-M) i_O
		\end{array}
		\right.
\end{equation}
with $\Gamma_E^2=j\omega (L'+M) \, j\omega C$ and $\Gamma_O^2=j\omega (L'-M) \, j\omega (C+2C_C)$. From these relations we deduce that, neglecting the loss in the resonator, for the even mode the effective inductance is $L_{eff,E}=L+M$ and the effective capacitance is $C_{eff,E}=C$. For the odd mode the inductance is $L_{eff,O}=L-M$ and the capacitance is $C_{eff,E}=C+2C_C$. For each section of the transmission line we have numerically, using radio-frequency simulation Sonnet software, computed the characteristic impedance $Z=\sqrt{L_{eff}/C_{eff}}$ and the velocity $v=1/\sqrt{L_{eff}C_{eff}}$ of the even and odd modes and extracted from them the value of the parameters of the resonant circuit (table \ref{CircuitConstants}). The values of the velocity for even and odd modes are very close to each other. It results the coupled lines must be seen as a single resonator where the resonant frequencies are only fixed by the impedance mismatch.  With those parameters it is possible to compute the impedance of the resonant circuit.

\begin{table}[htbp]
	\begin{center}
		\begin{tabular}{|c||c|c|c|c|c|c|c|c|}
		\hline
		Parameters & $L_a$ & $L_b$ & $M_a$ & $M_b$ \\
		\hline
		values & $1.047\mu H/m$ & $0.283\mu H/m$ & $0.615\mu H/m$ & $0.117\mu H/m$\\
		\hline
		\hline
		Parameters & $C_a$ & $C_b$ & $C_{C,a}$ & $C_{C,b}$  \\
		\hline
		values &  $43.7 pF/m$ & $182.5 pF/m$ & $61.1 pF/m$ & $125.8 pF/m$\\
		\hline
		\end{tabular}
	\end{center}
\caption{Values of the distributed inductances and capacitances of the resonator calculated from the geometry of the system (Fig.\ref{Circuit}).}
\label{CircuitConstants}
\end{table}

\subsubsection{Impedance matrix}

In the experiment, the junctions impedance can be changed in-situ by a dc voltage bias. Other parameters of the sample such as end resistances $R_{End}$, loss $R_{Loss}$ in the transmission line or source/detector coupling are fixed by the geometry and the choice of materials. The circuit can then be modelled by a $2\times2$ matrix which relates currents $i_S$ and $i_D$ to voltages $V_S$ and $V_D$ appearing at the source and detector stage. This matrix reads :

\begin{equation}
			\begin{pmatrix}
  		V_D  \\
  		V_S  \\
  		\end{pmatrix}=\overbrace{\begin{pmatrix}
  												Z_r & Z_t  \\
  												Z_t & Z_r   \\
  										\end{pmatrix}}^{\text{Impedance matrix}}.\begin{pmatrix}
  																				i_D  \\
  																				i_S  \\
  																	\end{pmatrix}
\label{Zt}
\end{equation}

where $Z_r$ is the complex impedance of the transmission lines resonator and $Z_t$ the transimpedance which quantifies the coupling between the junctions. The experiment allows to determine the real part of the impedance seen by the detector $Re[Z(\nu)]$. In the following, we describe how this can be done and highlight the link of this quantity with the elements of the matrix impedance. 

\begin{figure}[htbp]
  \begin{center}
			\includegraphics[width=8cm]{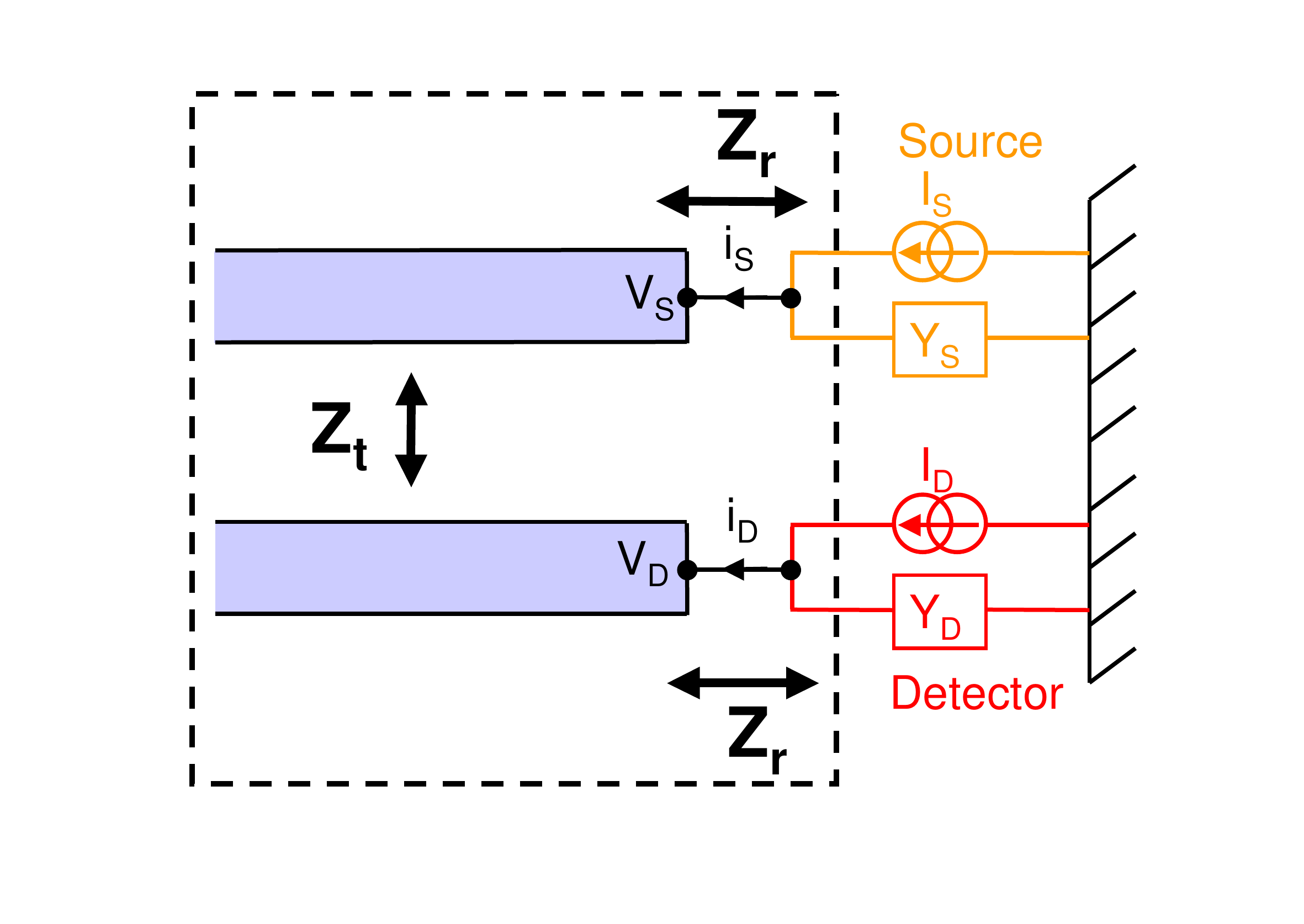}
	\end{center}
  \caption{Circuit used to model the source (admittance $Y_S$) and the detector (admittance $Y_D$) coupled by the resonator. It corresponds to a two channels HF circuit on which are connected the detector and the source junctions. This circuit is characterized by an impedance matrix (see text).}
  \label{CircuitPrinciple}
\end{figure}

\subsection{Experimental extraction of $Re[Z(\nu)]$ and relation with the matrix impedance}
\label{ReZextraction}

From now on and until the end of section \ref{QA}, the critical current of the detector junction is maximized (see Table \ref{SummaryConditions}). In this configuration, the $I(V)$ characteristic of the detector (small SIS junction) depends on the impedance of its electromagnetic environment \cite{ingold92}. We use this to calibrate the impedance seen by the detector. In the following paragraph the detector junction is voltage biased at low voltage, it can then be considered as an ac current source due to the ac Josephson effect. The source junction is not biased. 

\subsubsection{Principle}

As measured by Holst \textit{et al.} \cite{holst94}, in the particular case of a Josephson junction coupled to a superconducting transmission line resonator, current peaks in the I(V) of the junction appear in the subgap region $V_D<2\Delta/e$ due to the excitation of the resonator modes by the ac Josephson effect \cite{tinkham96,barone82}. These resonances in current are directly related to the resistive part of the impedance $Re[Z(\nu)]$ seen by the detector as:
\begin{equation}
	I(V)=\frac{I_C^2}{2}\frac{Re[Z(2eV/h)]}{V}=eI_C^2\frac{Re[Z(\nu)]}{h\nu} 
	\label{Ijj}
\end{equation}

with $I_C=\pi \Delta/(2eR_N)$ the critical current of the junction, $R_N=18.7k\Omega$ the normal state resistance of the junction \cite{tinkham96,ambegaokar63} and $\nu=2eV/h$, the Josephson frequency. This relation is deduced from the effect of the electromagnetic environment on the tunneling of Cooper pairs through the small Josephson junction \cite{ingold92}. 

\subsubsection{Measurement of $Re[Z(\nu)]$}

\begin{figure}[htbp]
  \begin{center}
		\includegraphics[width=8cm]{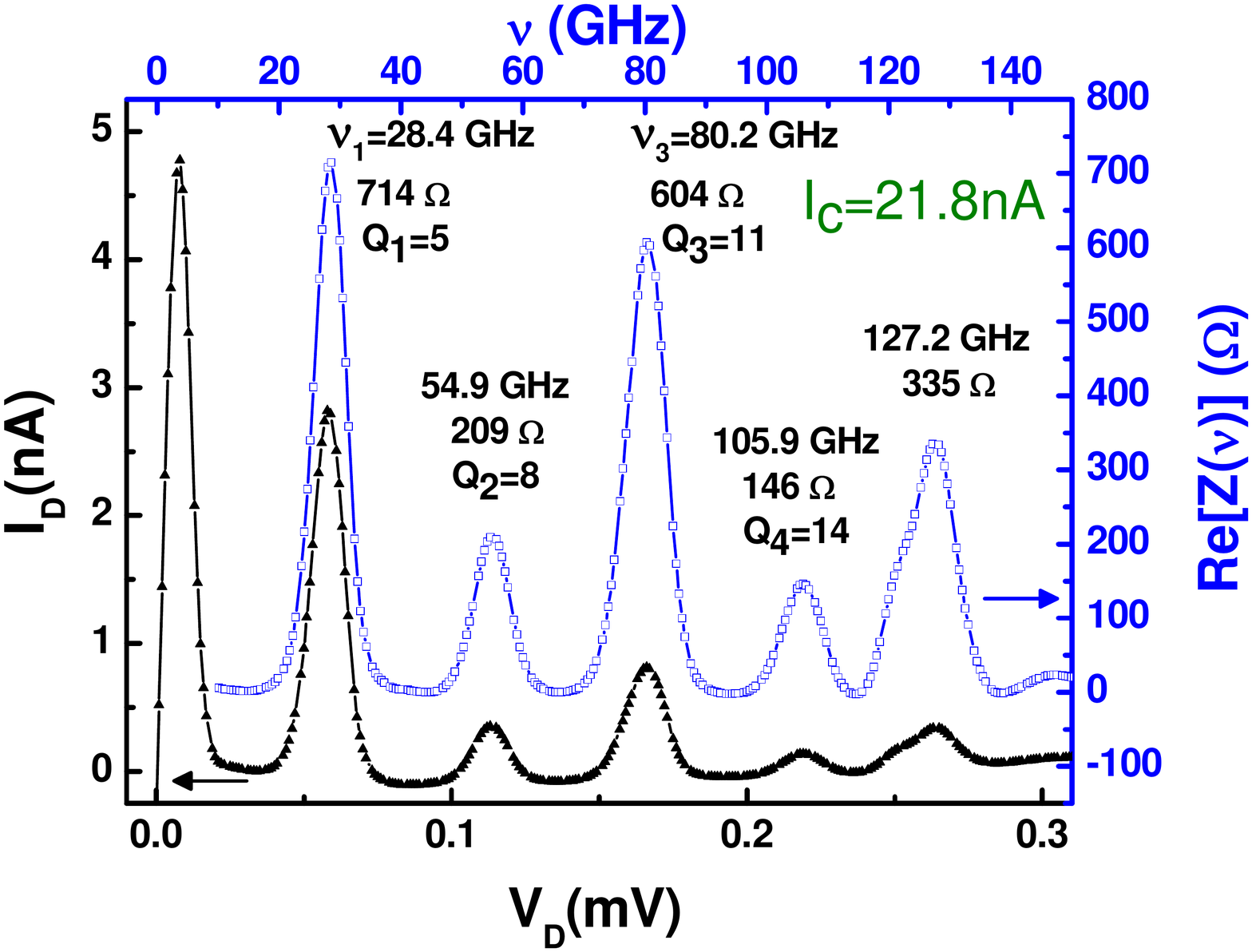}
	\end{center}
  \caption{Lower curve (left axis): experimental $I(V)$ dc characteristic of the detector junction in the subgap region with $I_C$ maximized by adjusting the magnetic flux. Upper curve (right axis) ~: the real part of the impedance seen by the detector junction, extracted from the previous curve using Eq. \ref{Ijj}, exhibits several resonances with quality factor $\approx$ 10.}
  \label{ReZseul}
\end{figure}

In figure \ref{ReZseul}, the $I(V)$ characteristic of the detector junction is shown in the subgap region for $I_C$ maximized with magnetic flux (see Table \ref{SummaryConditions}). Using Eq.\ref{Ijj} the subgap resonances allow to extract the real part of the impedance seen by the junction $Re[Z(\nu)]$. It exhibits peaks at frequencies $\nu_{1,2,3}=28.4,54.9$ and $80.2$GHz. With a length $l=1$mm the first resonance was expected at 30 GHz. We attribute the difference with the measured resonant frequency to the capacitance of the junction which shifts the resonance. The relatively low value of the quality factors $Q_n$ will be discussed in the next section.
The fact that we see resonances at frequencies $\nu_n=nv/4l$, with $n$ not only an odd integer but also an even integer is attributed to the rather small ratio $Z_a/Z_b < 10$ of the impedances of the transmission lines. The amplitude of the peak at $28GHz$ is equal to $714\Omega$ in agreement with condition of validity of Eq.\ref{Ijj} \textit{i.e.} $Re[Z]<<9400\Omega$\footnote{Relation \ref{Ijj} is valid when $E_J \, P'(E)<<1$. We compute $E_J=\frac{\hbar I_C}{2e}\approx40\mu eV$. At $2eV=h\nu_1$ with $\nu_1 \approx 28 GHz$, one obtains that $Re[Z(2eV/h)]<<\frac{h\nu_1 R_Q}{2E_J}\approx9400\Omega$ is a sufficient condition for this expression to be valid.}. 

\subsubsection{Relation between $Re[Z(\nu)]$ and $Z_r(\nu)$ in the matrix impedance}
\label{ReZmeasVsZres}

In this part we address the relation between the measured quantity $Re[Z(\nu)]$ and the impedance $Z_r(\nu)$ of the resonator entering in the matrix impedance (Eq.\ref{Zt}). Approximating first the detector junction as a ac Josephson current source of negligible admittance we find that the presence of the impedance $Y_S$ at the end of line is responsible for a renormalization of the impedance of the resonator: 
\begin{equation}
 Z_{r,eff}(\nu,Y_S)=Z_r(\nu)-\frac{Y_S(\nu,V_S)Z_t(\nu)^2}{1+Y_S(\nu,V_S)Z_r(\nu)}.
\end{equation} 
Moreover, the detector does not behave as a pure ac Josephson current source and one also has to consider its finite admittance $Y_D$. This leads to a measured $Z(\nu)$ which reads (see figure \ref{CircuitPrinciple}):
\begin{equation}
Z(\nu)=\frac{Z_{r,eff}(\nu,Y_S)}{1+Y_D(\nu)Z_{r,eff}(\nu,Y_S)}
\label{ZFINAL}
\end{equation}

\begin{figure}[htbp]
  \begin{center}
			\includegraphics[width=8cm]{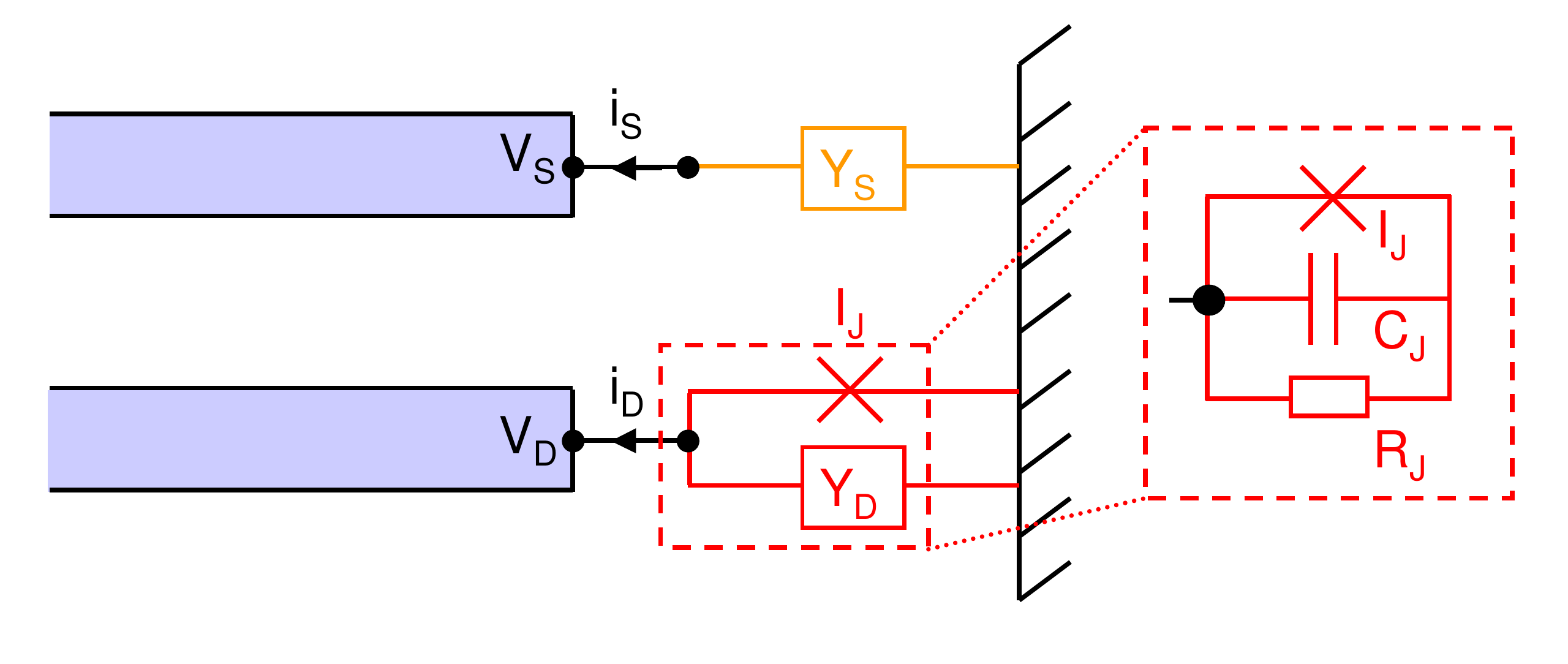}
	\end{center}
  \caption{Modeling of the ac currents and voltages through and across the resonator in the $Re[Z(\nu)]$ measurement. The measured impedance seen by the detector depends on two parameters which are the admittance of the detector $Y_D$ ($RC$ parallel circuit) and the effective impedance of the resonator $Z_{r,eff}$. This latter quantity does itself depend on $Z_r$ and the source impedance $Y_S$ via the transimpedance $Z_t$ (see text).}
  \label{Circuit2}
  \end{figure} 

When computing $Z$ for small real admittances $Y_D(\nu)$ and $Y_S(\nu)$ ($Y_{D,S}<<1/Z_r,1/Z_t$ which is the case in the experiment) one straightforwardly finds that $Z\approx Z_r$. On the other hand when $Y_S(\nu)$ and $Y_D(\nu)$ are pure capacitances $Y_S(\nu)=Y_D(\nu)=jC_{J}2\pi\nu$, $Z$ exhibits resonances shifted compared to the original $Z_r$ with identical amplitudes and quality factors. In general, a precise analysis of $Z$ yields information on the impedance of the source. This is exploited in section \ref{QA}.
  
\subsection{Comparison with the experiment}

\begin{figure}[htbp]
  \begin{center}
			\includegraphics[width=8cm]{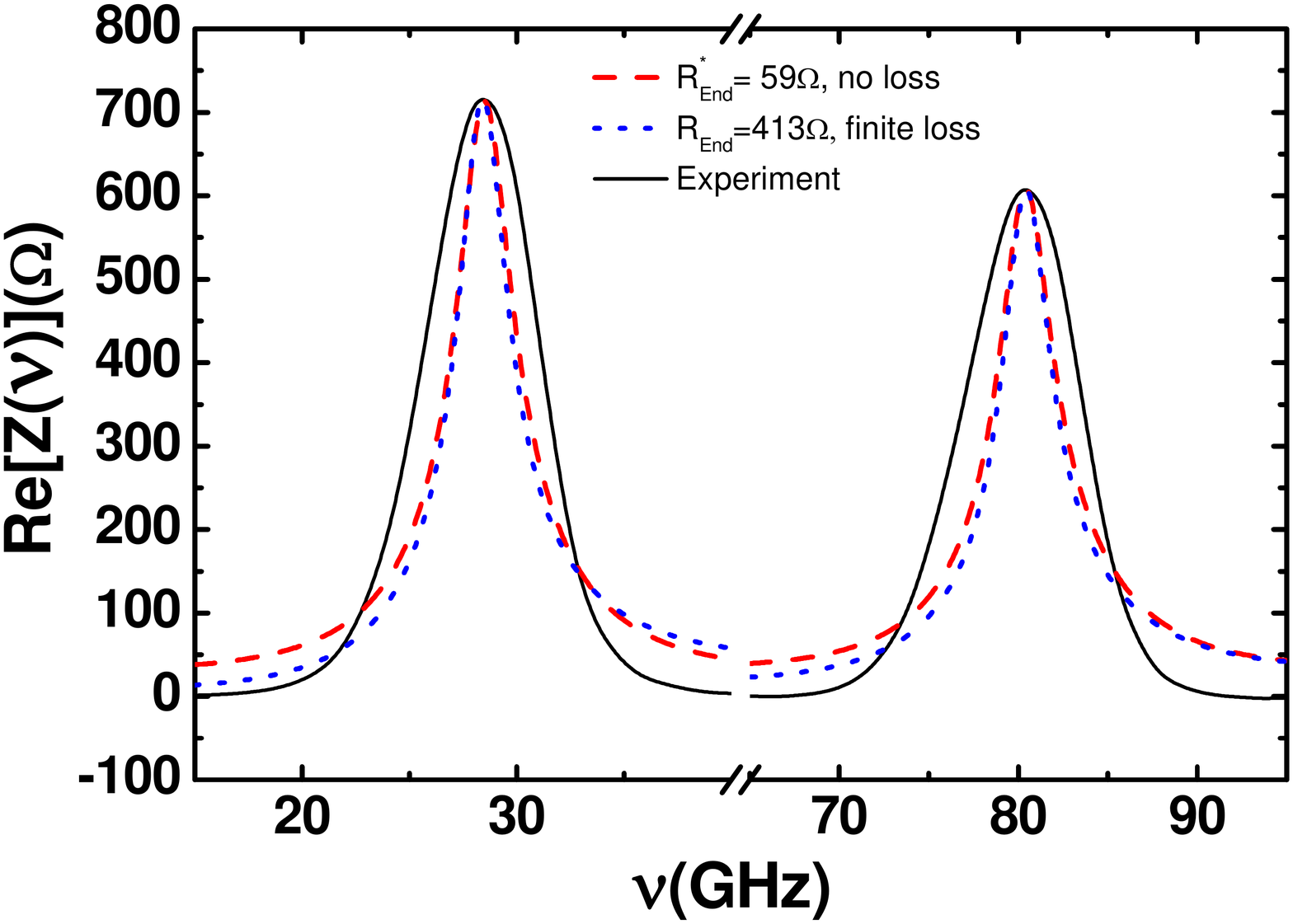}
	\end{center}
  \caption{Black full lines : experimental $Re[Z(\nu)]$ seen by the detector. Red dashed curves : calculated curves obtained without dissipation, i.e. $R_{Loss}=\infty$, and an effective $R^*_{End}$ ($R^*_{End}=59\Omega$ near $\nu_1$ and $R^*_{End}=50.5\Omega$ near $\nu_3$ ). Blue dotted curves : calculated curves obtained by considering $R_{End}=413\Omega$ with finite dissipation ($R_{Loss}=1.36M\Omega/m$ near $\nu_1$ and $R_{Loss}=10.15M\Omega/m$ near $\nu_3$ ).}
  \label{ReZfit}
\end{figure}

We now propose to fit the experimental curve of $Re[Z(\nu)]$ with the model previously described (Fig.\ref{ReZfit}). This is done in two ways.

First we fix $R_{End}=413\Omega$. The value of the detector resistance and capacitance (Fig \ref{Circuit2}) is set to $R_J=10M\Omega$ and $C_J=7fF$, their expected values. With these parameters, it is necessary to introduce losses along the transmission line $R_{Loss}(\nu_1)\approx1.36M\Omega/m$, and $R_{Loss}(\nu_3)\approx10,15M\Omega/m$ to reproduce the amplitude of the peaks. At the same time the quality factors are approximately $2$ times higher than in the experiment. These losses are unphysical because aluminum and the low doping substrate used in the experiment are not known for exhibiting such a large amount of losses \cite{wang09}.

Second we fit the experimental data by taking an effective $R^*_{End}$ without intrinsic losses. The fit reproduces the peak amplitude with $R^*_{End}=59\Omega$ near $\nu_1$ and $R^*_{End}=50.5\Omega$ near $\nu_3$. The quality factor discrepancy is still of a factor $\approx 2$. This points out the difficulty to take into account all the dissipation sources in our model. The effective value of the end resistor is smaller than expected ($R^*_{End}<<R_{End}$) due to dc polarization lines capacitively coupled to the ground.

\subsection{Conclusion}

We have modelled the resonant coupling circuit and compared our results to experimental data. It allowed us to qualitatively understand the role of several sample parameters on the efficiency of the coupling circuit. The impedance at the end of the resonator is found to be a critical parameter.

\section{High Frequency Admittance Measurement with an On-chip Resonant Circuit}
\label{QA}

A relevant quantity characterizing the quantum dynamics of mesoscopic systems is the high frequency admittance. However, whereas its measurement is of great interest, only few experiments managed to measure this quantity \cite{worsham91,hu90,tucker85} due to the difficulty to work at high frequency. 
In this part, we show how, using the  setup described previously, one can measure the bias dependence of the complex admittance of the source $Y_S=Re[Y_S]+iIm[Y_S]$ at the resonant frequencies of the resonator.  In the following, $Re[Y_S]$ is called the quantum conductance and $Im[Y_S]$ the quantum susceptance of the source \cite{hu90,tucker85,worsham91} \footnote{Note than in order to verify the generalized fluctuation dissipation theorem, one needs to access the dissipative component of the admittance.}. 

\subsection{Principle of the experiment}
\label{MPQA}
We have derived in section \ref{ReZextraction} the relation \ref{ZFINAL} between the impedance seen by the detector and the admittance of the source junction $Y_S$. By measuring the dependence of $Re[Z(\nu)]$ \textit{vs} the bias voltage $V_S$ of the source junction we can determine $Y_S(V_S)$. Hereafter we will assume that the measured impedance $Z(\nu)$ is identical to $Z_r(\nu)$ shifted by a quantity proportional to the geometrical capacitances of the junctions (source and detector). The sensitivity of the experiment is calculated by using equation \ref{ZFINAL} and by assuming the resonant peaks of $Z(\nu)$ and $Z_t(\nu)$ extracted from the experiment can be approximated  by a sum of Lorentzian functions centered around resonant frequencies of the resonator so that:
\begin{eqnarray}
	Z(\nu)&=&\sum^{n}_{i=0} \frac{Z_{i}}{1+j\alpha_i(\nu-\nu_i)}\\
	Z_t(\nu)&=&\sum^{n}_{i=0} \frac{Z_{t,i}}{1+j\beta_i(\nu-\nu_i)}
\end{eqnarray}
with $\nu_i$ the $i^{th}$ eigen frequency, $Z_{i}$ and $Z_{t,i}$ the amplitude of the resonance and $\alpha_i$ and $\beta_i$ fitting parameters corresponding respectively to the width of the resonances of $Z(\nu)$ and $Z_t(\nu)$ at frequency $\nu_i$.

\begin{figure}[htbp]
  \begin{center}
		\includegraphics[width=8cm]{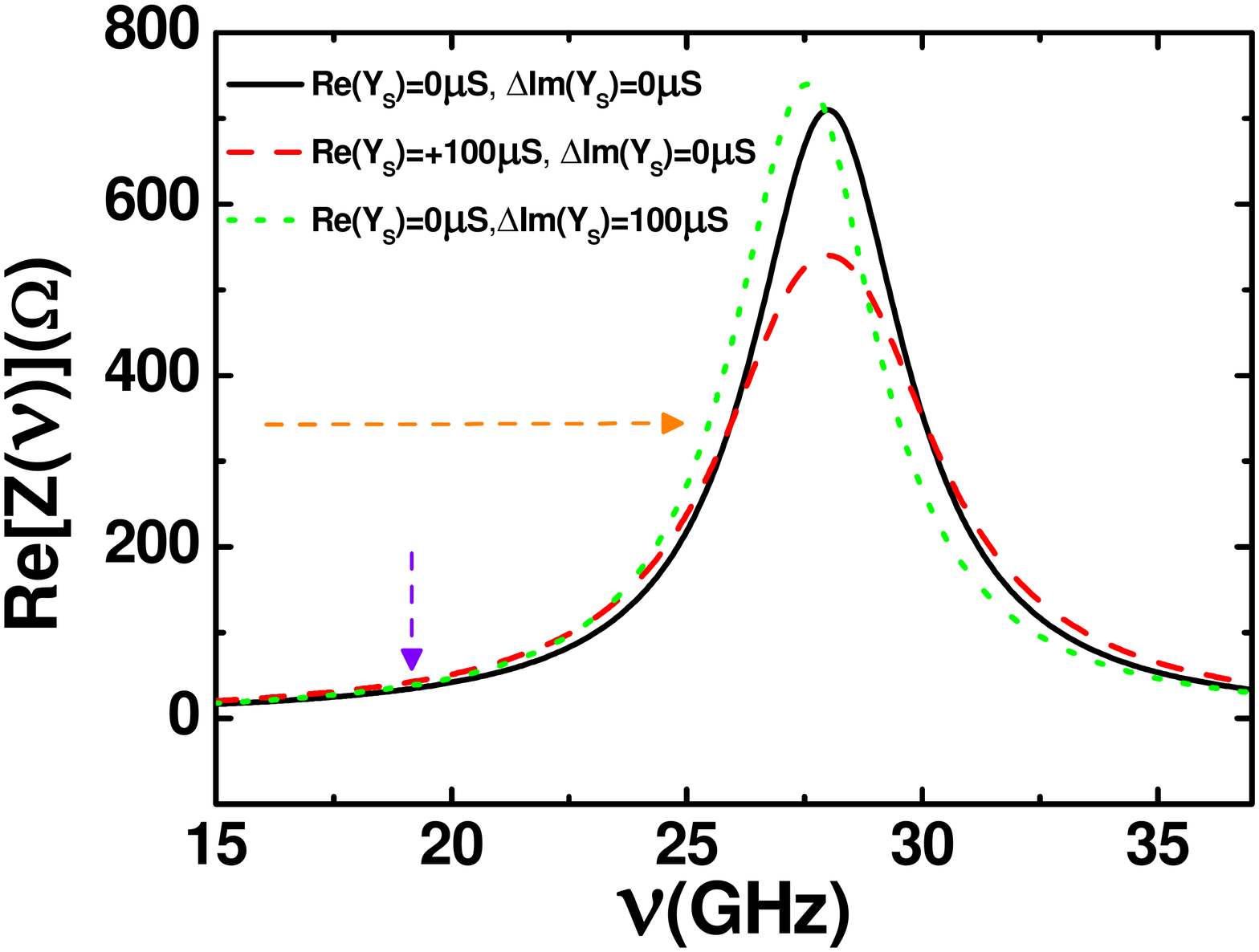}
	\end{center}
  \caption{Calculated evolution of $Re[Z(\nu)]$ as a function of the real $Re[Y_S]$ and imaginary $Im[Y_S]$ components of the source junction impedance near $\nu_1$. The calculations are done by using the best Lorentzian fits of the experimental curves of $Re[Z]$ and $|Z_t|$. The specific point denoted by the horizontal dashed arrow is not affected by $Re[Y_S]$ changes while it is very sensitive to changes of $Im[Y_S]$. Even far from resonance, influence of $Re[Y_S]$ on quality factors is visible as shown by the dashed vertical arrow.}
  \label{ReZ=f(ConductY2)}
\end{figure}

We present in figure \ref{ReZ=f(ConductY2)} some calculated curves of $Re[Z(\nu)]$  for different values of the admittance $Y_S$ around $\nu_1$. As expected \cite{deblock02}, an increase of $Re[Y_S]$ ($Im(Y_S)$ is fixed to its geometrical value $jC_{J}2\pi\nu$) reduces the amplitude and the quality factor of the resonance without changing the resonant frequency. This behaviour is roughly linear with respect to a moderate increase of $Re[Y_S]$ and the inflection point of the Lorentzian peak, denoted by the horizontal dashed arrow, is not affected by $Re[Y_S]$ changes. On the other hand, when one computes the same curve for a different imaginary part of the admittance of $Y_S$ ($Re[Y_S]$ is fixed to $0$, $\Delta Im[Y_S]>0$) the resonant frequency is shifted at a lower frequency proportionally to $\Delta Im[Y_S]$. At the same time, both amplitudes and quality factors are nearly unchanged.

\subsection{Sensitivity of the detector to the high frequency admittance of the source}
From the previous subsection, we conclude that by measuring the shape and position of the resonances  we can access the bias dependence of the real and imaginary part of $Y_S$. This is exploited in the following using the detector in the ac Josephson regime as a generator like in section \ref{ReZextraction}. In the experiment the quality factors are low, the amplitude of the peaks are weak and the voltage biasing is not perfect. To increase sensitivity we ac bias the source and measure the resulting signal on the detector with a lock-in amplifier technique. 
In the following we compare two different techniques we used to determine $Re$ and $Im[Y_S]$ as a function of $V_S$ by using the detector either in a voltage or a current bias mode. In both cases we modulate the source voltage $V_S$ and measure by lockin detection either the induced current modulation $\partial I_D/\partial V_S$ or voltage modulation $\partial V_D/\partial V_S$  on the detector (see Table \ref{SummaryConditions}).

\subsubsection{Voltage biased detector}
\begin{figure}[H]
  \begin{center}
	 \includegraphics[width=8cm]{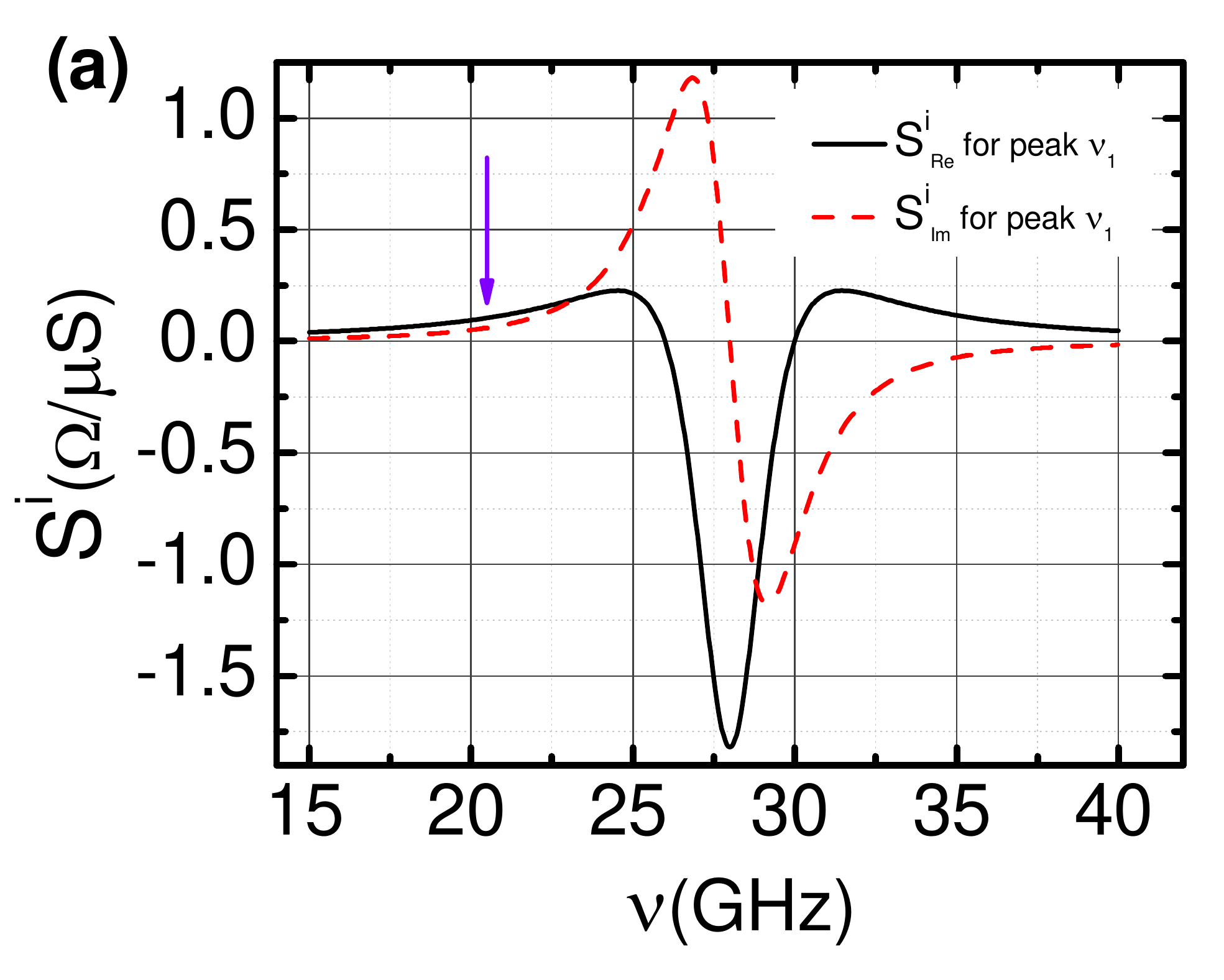}
	 \includegraphics[width=8cm]{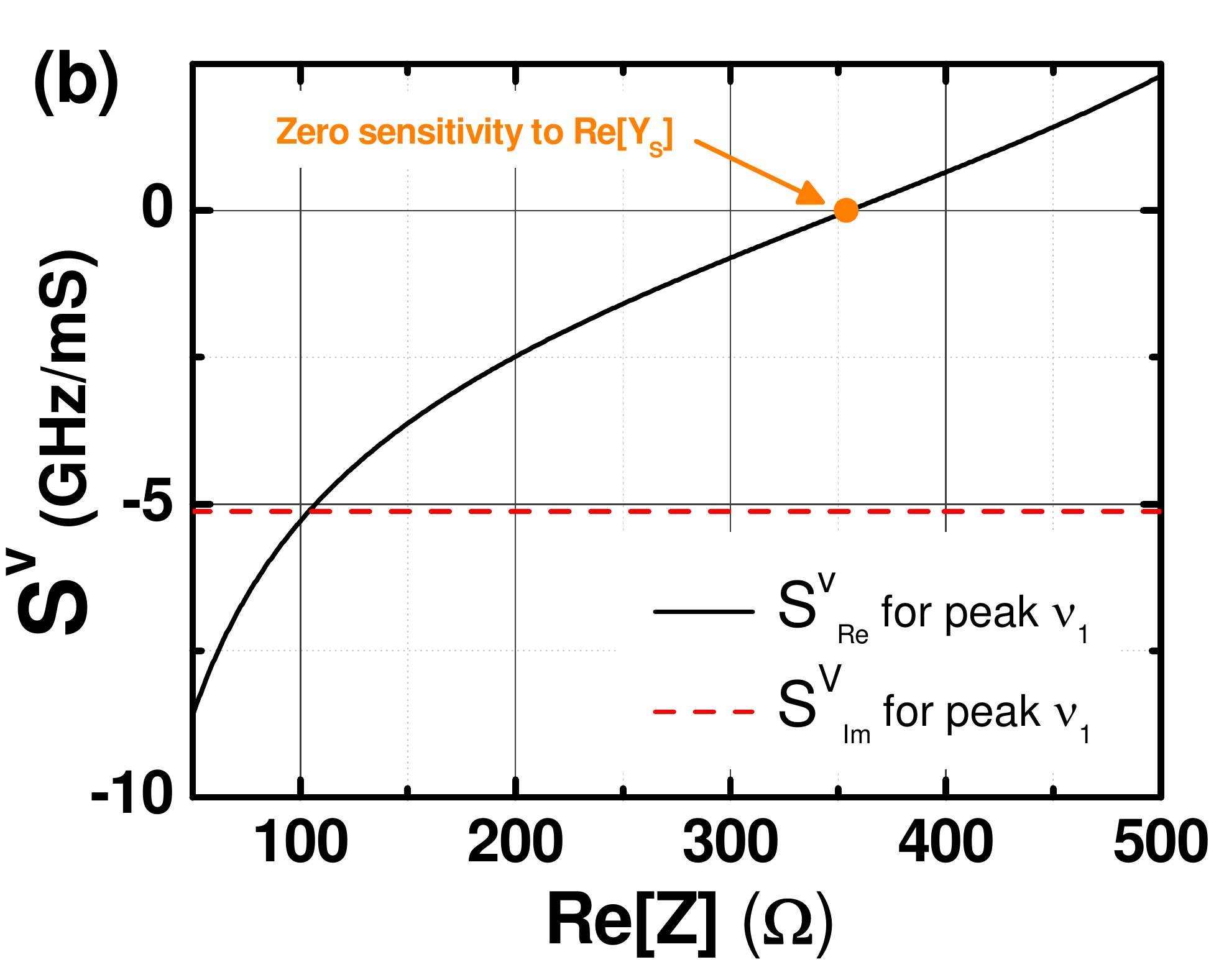}
	\end{center}
  \caption{(a) Calculated sensitivity  of $Re[Z]$ to the real ($S^i_{Re}$) and imaginary part ($S^i_{Im}$) of the admittance of the source $Y_S$ near $\nu_1$ for the voltage biased detection scheme. The sensitivity to the imaginary part $S^i_{Im}$ is either positive or negative depending on which side of the resonance the measurement is done. It is zero at the resonant frequency. Opposingly, the sensitivity to the real part $S^i_{Re}$ is maximum at the resonant frequency. Far from resonance, the vertical dashed arrows denotes the chosen detector position where $S^i_{Re}$ is greater than $S^i_{Im}$. (b) Calculated sensitivity  of $Re[Z]$ to the real ($S^v_{Re}$) and imaginary part ($S^v_{Im}$) of the admittance of the source $Y_S$ near $\nu_1$ for the current biased detection scheme. $S^v_{Im}$ is almost constant along the  resonance peak whereas $S^v_{Re}$ is zero at the point denoted by the arrow. At this point the detector is thus only sensitive to $Im[Y_S]$. The same predictions can be done near $\nu_3$.}
  \label{Condsensitivity}
\end{figure}

We describe the sensitivity of the detector to a variation of  $Re$ or $Im[Y_S]$ by 2 coefficients: $S^i_{Re}$ and $S^i_{Im}$ respectively equal to  $\partial Re[Z]/\partial Re[Y_S]$ and $\partial Re[Z]/\partial Im[Y_S]$. The current response of the detector to a low frequency modulation of $V_S$ is then:
\begin{equation}
\frac{\partial I_D}{\partial V_S} = \frac{\partial I_D}{\partial Re[Z]} \left[S^i_{Re}\frac{\partial Re[Y_S]}{\partial V_S}+ S^i_{Im}\frac{\partial Im[Y_S]}{\partial V_S}\right]
\end{equation}
 $\partial I_D/\partial Re[Z]$ can be calculated using Eq.\ref{Ijj} and is equal to $eI_C^2/h\nu$ (see part \ref{ReZextraction}) with $\nu$ the Josephson frequency $2eV_D/h$ associated with the bias of the detector. The quantities $S^i_{Im}, S^i_{Re}$ can be numerically computed and are plotted in fig.\ref{Condsensitivity}  as a function of the frequency in the vicinity of the fundamental mode of the resonator. As expected $S^i_{Re}$ is maximum at resonance whereas $S^i_{Im}$ is  equal to zero. On the other hand $S^i_{Im}$ is maximum at the inflexion point of the resonance. Unfortunately the stability of the detector is not excellent and it was not possible to adjust precisely the detection frequency at these optimum points. A better accuracy was obtained  on the detection of $\partial Re[Y_S]/\partial V_S$ when the detector was polarized below the inflexion point of the resonance denoted by vertical dashed arrows (bottom of the peaks) in figure \ref{Condsensitivity} a. For this bias value of the detector, the contribution of $Im[Y_S]$ is  lower than the contribution of $Re[Y_S]$ by a factor $0.56$.

\subsubsection{Current biased detector}
  A better sensitivity  for the determination of $\partial Im[Y_S]/\partial V_S$ is obtained   by \textit{current}   instead of voltage biasing the detector and measuring voltage.  The current  value is adjusted on the side of the investigated  resonance peak (taking advantage of the hysteresis behavior of $V_D(I_D)$). The quantity   measured is then the derivative of the dc voltage across the detector versus the bias voltage of the source $\partial V_D/\partial V_S$ with a lock-in detector. It is related to the admittance of the source as follows : 
 \begin{equation}
\frac{\partial V_D}{\partial V_S} = \frac{2e}{h}  \left[S^v_{Re}\frac{\partial Re[Y_S]}{\partial V_S}+ S^v_{Im}\frac{\partial Im[Y_S]}{\partial V_S}\right]
\end{equation} 
  
 Where  the quantities $S^v _{Re} =\partial \nu /\partial Re[Y_S]$, and $S^v _{Im} =\partial \nu /\partial  Im[Y_S]$ describe the voltage sensitivity of  the detector (in frequency units) to a variation of $Re$ and $Im[Y_S]$. They   are depicted in Fig.\ref{Condsensitivity}b as a function of the current bias on the detector.

From the analysis of these curves, we conclude that by current biasing the detector at the inflection point of the resonant peaks we measure a signal only proportional to $Im[Y_S]$, the imaginary part of the source junction. This position is denoted by  the arrows in figure \ref{Condsensitivity}b.

\subsection{High frequency admittance of the source Josephson junction}

\begin{figure}[htbp]
  \begin{center}
		\includegraphics[width=8cm]{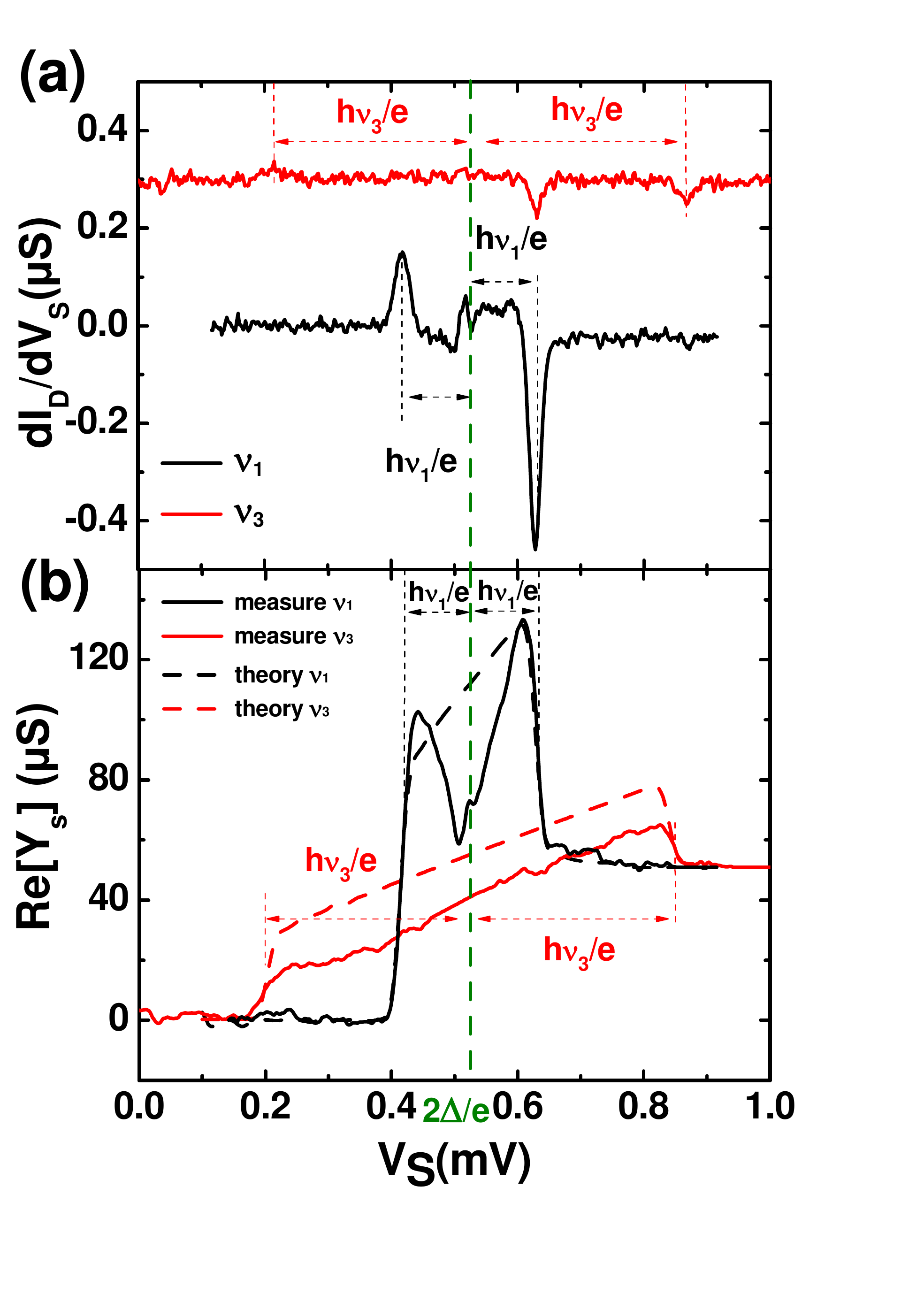}
	\end{center}
  \caption{(a) Experimental $dI_D/dV_S$ curves obtained by voltage biasing the detector at the bottom of the resonant peaks near $\nu_1$ and $\nu_3$ and measuring the detector current modulation versus the voltage bias $V_S$ of the source (see text). The curves are vertically shifted for the sake of clarity. (b) The solid line curves are the experimental quantum conductance data, near the quasiparticles branch, extracted from curves shown in (a). It is obtained after removing the emission noise contribution and numerical integration with respect to the source bias voltage $V_S$, taking into account the calculated sensitivity. The dashed lines correspond to the quantum conductance predicted by theory.}
  \label{ReG}
\end{figure}

In the following we show how one can deduce the full bias dependence of the complex admittance of the source junction \cite{worsham91} at the eigen frequencies of the resonator.
There are two contributions to the measured signals. Beside the contribution of the bias dependent admittance of the source we have to consider the finite frequency shot noise contribution which can be measured independently (see section \ref{NoiseJoseph}) and thus removed from the data
\footnote{To remove the noise contribution we assume that for $V_S >> 2\Delta/e$, the quantum admittance is constant and thus its derivative is zero. The noise, however leads to a constant value in the measured signal. We can thus subtract the contribution of the emission noise.}. We finally integrate the signal function with respect to $V_S$ and deduce  $Re$ and $Im[Y_S]$ using the coefficients $S^i$ and $S^v$ calculated previously. The bias dependent  quantum conductance $Re[Y_S(V_S)]$ and susceptance $Im[Y_S(V_S)]$ of the source Josephson junction are thus measured at the two frequencies $\nu_1$ and $\nu_3$ depending on which peak the detector is biased (see figure \ref{ReG} b). It can be compared to the theoretical expressions derived in \cite{tucker85}: 
\begin{eqnarray}
\nonumber Re[Y_S(V_S)]&=&\frac{e}{2h\nu}[I_{QP}(V_S+nh\nu/e)-I_{QP}(V_S-nh\nu/e)]\\
\nonumber Im[Y_S(V_S)]&=&\frac{e}{2h\nu}[I_{KK}(V_S+nh\nu/e)+I_{KK}(V_S-nh\nu/e)\\
&-&2I_{KK}(V_S)]
\label{GQ}
\end{eqnarray}
where $I_{QP}(V_S)$ is the I(V) characteristic of the junction close to the gap and $I_{KK}(V_S)$ the Kramers-Kronig transform of $I_{QP}(V_S)$ , defined as : 
\begin{equation}
I_{KK}(V_S)=P \int^{\infty}_{-\infty}\frac{dV_S'}{\pi}\frac{I_{QP}(V_S')-V_S'/R_n}{V_S'-V_S}.
\end{equation}

 $R_n$ is the normal state resistance of the junction and $P$ represents the Cauchy principal value.

We plot in figure \ref{ReG}b, the theoretical and experimental curves obtained for the measured Josephson junction. 

Concerning the quantum conductance $Re[Y_S]$, the result of the experiment is accurate concerning the positions of the predicted steps for both measurement frequencies $\nu_1$ and $\nu_3$. On the other hand the experimental data exhibit an additional dip in the interval $[2\Delta/e-h\nu_1/e,2\Delta/e+h\nu_1/e]$. We attribute this difference with theoretical prediction to a direct cross talk  between the source junction and the detector in $V_S=2\Delta/e$ and to a non-zero sensitivity to the imaginary part of the junction admittance. For frequency $\nu_3$, the detection is less sensitive and the cross talk in $V_S=2\Delta/e$ is not visible anymore. The quantum conductance extracted is thus more reliable in the interval $[2\Delta/e-h\nu_3/e,2\Delta/e+h\nu_3/e]$ than for $\nu_1$. However, its experimental amplitude is approximately $2$ times smaller than the expected one. We attribute this difference to a lack of sensitivity of the detector at this frequency. 

\begin{figure}[htbp]
  \begin{center}
		\includegraphics[width=8cm]{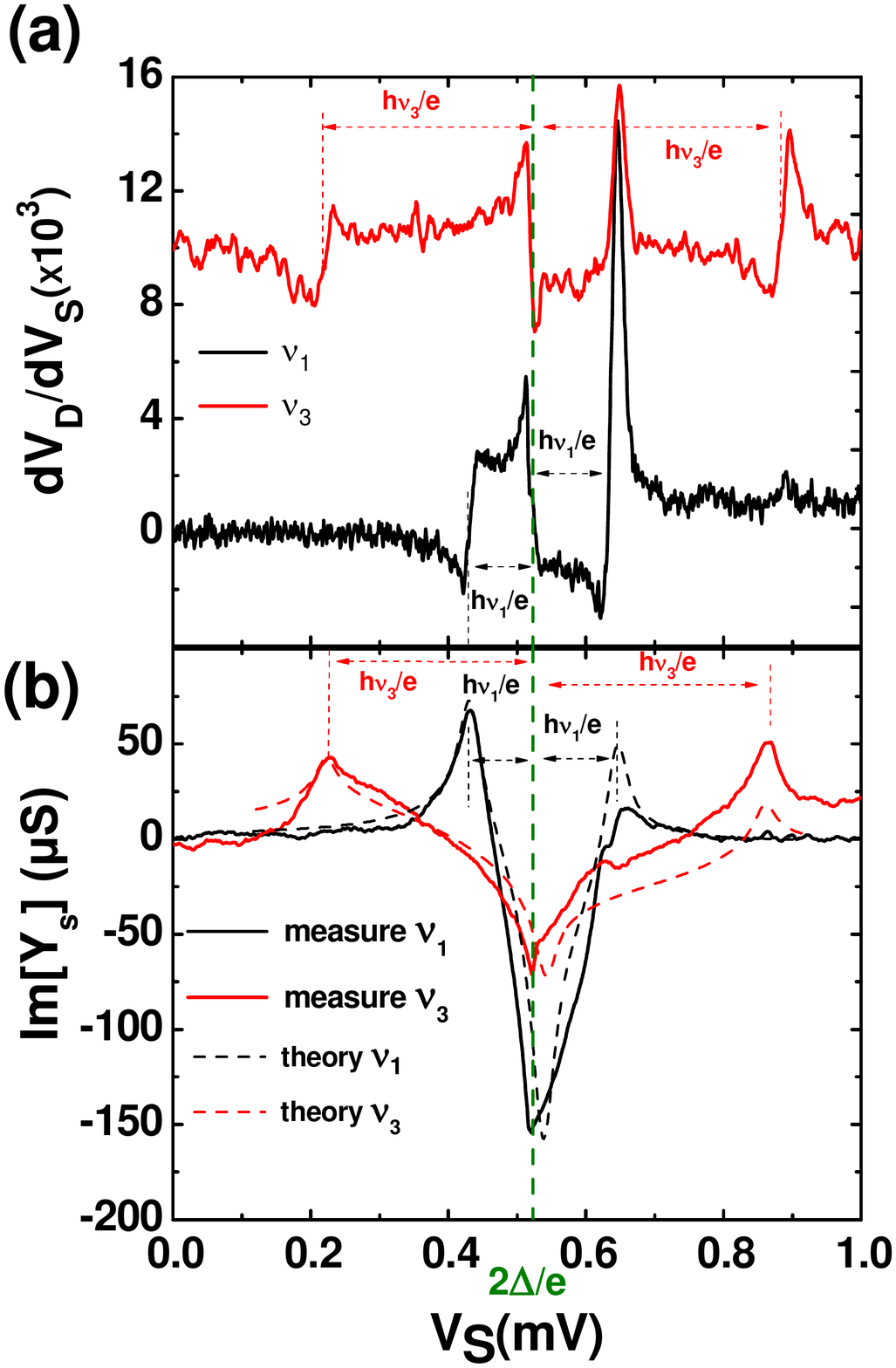}
	\end{center}
  \caption{(a) Experimental $dV_D/dV_S$ curves obtained by current biasing the detector at the middle edge of the resonant peaks $\nu_1$ and $\nu_3$ and measuring the detector voltage modulation versus the voltage bias of the source. The curves are vertically shifted for the sake of clarity. (b) Experimental (solid lines) quantum susceptance, near the quasiparticles branch, obtained after removing the emission noise contribution and integrating numerically the data with respect to the source bias voltage $V_S$, taking into account the calculated sensitivity. The dashed lines correspond to the theory.}
  \label{fig10}
\end{figure}

We also obtained an experimental determination of the quantum susceptance $Im[Y_S]$ (see Fig.\ref{fig10}). A reasonable agreement between theory and experiment is found, both regarding the amplitude of the signal and the positions of the predicted singularities.

\subsection{Conclusion}
We have presented a semi-quantitative way to measure the high frequency admittance of a high impedance nanodevice, the Josephson junction. The obtained results, even though they reasonably agree with theoretical predictions, are not as precise as previous measurements reported in references \cite{hu90,worsham91}. However we can use the same setup to measure the out-of-equilibrium quantum noise of the source under interest. This is what is discussed in the following part.

\section{SIS junction as a high frequency noise detector}
\label{SISNoiseDetector}

To measure nonsymmetrized noise, i.e., distinguish between emission and absorption, one can use a quantum detector \cite{clerk10,aguado00}. Different realizations of such a detection scheme have been implemented by using, e.g., quantum bits \cite{schoelkopf02}, quantum dots \cite{onac06,gustavsson07}, a superconducting resonator \cite{xue09} or a superconductor-insulator-superconductor (SIS) tunnel junction \cite{deblock03,billangeon06,basset10}.
Here we develop the case of the SIS junction which will be used as a quantum detector of noise able to distinguish the emission and absorption parts of the noise when looking at inelastic tunneling of quasiparticles. We emphasize this point explaining the principle of detection. From now on, the critical current of the detector is minimized and we are only considering its quasiparticles current (see Table \ref{SummaryConditions}). 

\subsection{Photo-assisted quasiparticles tunnelling current as a probe of voltage noise}

The dependence of quasiparticles tunneling in SIS junction versus irradiation with microwave photons has been widely used for making mixers \cite{tucker85}. More recently SIS junctions have been used as quantum detectors of noise in mesoscopic physics \cite{deblock03,billangeon06}. The way in which the $I(V)$ characteristic of such a detector is modified is understood in terms of photo-assisted tunneling (PAT) current. This current adds to the elastic current in presence of voltage fluctuations accross the junction. 
In such a configuration the total quasiparticles current which flows through the detector can be written as \cite{ingold92}:

\begin{equation}
I_{QP}(V_D)=\int^{+\infty}_{-\infty} dE' \; \frac{1-e^{-\beta eV_D}}{1-e^{-\beta E'}} P(eV_D-E') I_{QP,0}(\frac{E'}{e})
\label{IQPtotThwithBeta}
\end{equation}

with $P(E)$ the probability to exchange the energy E with the environment during a tunneling process, $I_{QP,0}(V_D)$ the $IV$ characteristic of the junction without environment and $\beta=1/k_BT$. Note than this treatment only holds for small capacitance and high normal state resistance of the junctions.

In the limit of small noise power, the relation between the probability $P(E)$ and the voltage fluctuations $S_V(\nu)$ of the environment can be expressed, at low temperature \cite{aguado00}, as:

\begin{equation}
P(E) = \left[1-\frac{e^2}{h^2}\int^{+\infty}_{-\infty} d\nu \frac{S_V(\nu)}{\nu^2} \right]\delta(E)+\frac{e^2}{h}\frac{S_V(E/h)}{E^2}.
\label{LastPdeE}
\end{equation}

From expressions \ref{LastPdeE} we obtain the PAT current through the detector as a function of its bias voltage $V_D$ and the non-symmetrized spectral density of voltage noise $S_V(\nu)$ across the detector junction \cite{ingold92,aguado00} which reads, in the limit $k_B T \ll eV_D$ and small voltage noise $S_V(\nu)$ :  
\begin{eqnarray}
	\nonumber I_{PAT}(V_D)&=&I_{QP}(V_D)-I_{QP,0}(V_D)\\
	\nonumber &=&\int^{\infty}_{0} d\nu \left( \frac{e}{h\nu} \right)^2 S_V(-\nu)I_{QP,0}(V_D+\frac{h\nu}{e})\\
	\nonumber &+&\int^{eV_D/h}_{0} d\nu \left( \frac{e}{h\nu} \right)^2 S_V(\nu)I_{QP,0}(V_D-\frac{h\nu}{e})\\
 	&-&\int^{+\infty}_{-\infty} d\nu \left( \frac{e}{h\nu} \right)^2 S_V(\nu)I_{QP,0}(V_D)
	\label{IPAT}	
\end{eqnarray}
with $I_{QP,0}(V_D)$ the $I(V)$ characteristic of the detector without electromagnetic environment. The first term of Eq.\ref{IPAT} is related to the emission noise, the second to the absorption noise and the third corresponds to the elastic current. 

\begin{figure}[htbp]
  \begin{center}
			\includegraphics[width=8cm]{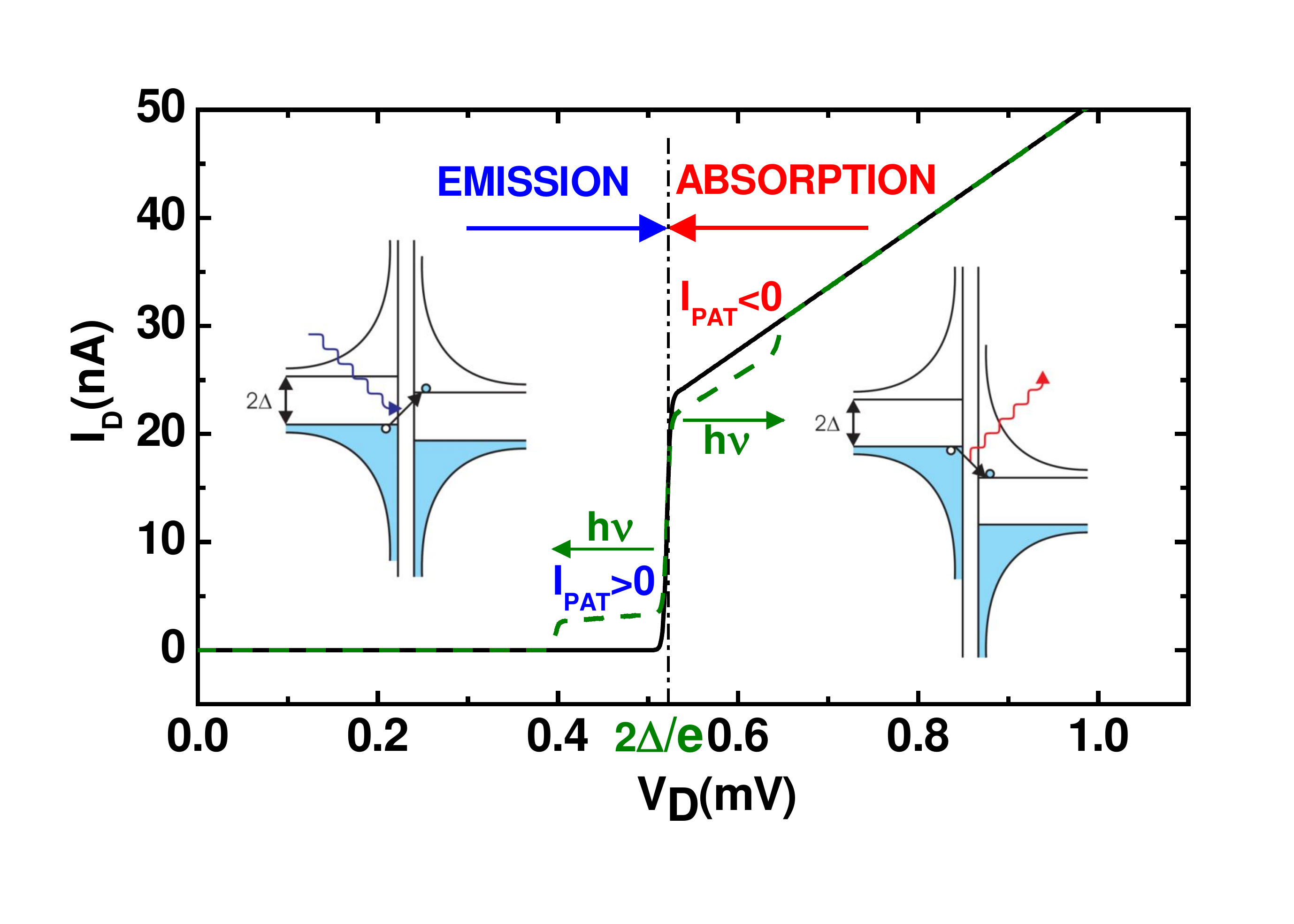}
	\end{center}
  \caption{Solid black curve: Current voltage characteristic of a typical SIS junction. Dashed green curve : calculated IV characteristic of the detector under external high frequency irradiation giving rise to photo-assisted tunnelling current $I_{PAT}$. Below the gap, the detector is \textit{emission} sensitive. Above the gap, the detector is mainly \textit{absorption} sensitive. The amplitude of the PAT current is arbitrarily increased for the purpose of clarity. Schematic drawings correspond to semiconducting representations of the SIS junction when it is voltage biased below or above the gap. Below the gap a photon emitted by the environment allows the tunneling of a quasiparticle from one superconductor to the other (positive PAT current). Above the gap a photon absorbed by the environment decreases the overall tunneling rate of quasiparticles from one superconductor to the other (negative PAT current).}
  \label{IVSemiconRep}
\end{figure}

\subsection{Emission noise sensitive region}
\label{EmissNoisRegion}
When $|V_D|<2 \Delta/e$, due to the superconducting density of states, $I_{QP,0}(V_D)=0$. Expression \ref{IPAT} shows that only emission noise is detected for frequencies higher than $(2 \Delta-e V_D)/h$. This region is emission noise sensitive (see Fig.\ref{IVSemiconRep}). The PAT current through the junction is positive due to the energy coming from the environment (emitted noise) which increases the net current flowing through the detector.

\subsection{Absorption noise sensitive region}
\label{AbsNoisRegion}
From formula \ref{IPAT}, we deduce that for $|V_D| > 2 \Delta/e$  the detector is mainly sensitive to absorption noise. This region is absorption noise sensitive (see Fig.\ref{IVSemiconRep}). The PAT current which adds to the large positive dc current is negative. The energy lost in the environment (absorbed noise) reduces the net current flowing through the detector.

\section{Out-of-Equilibrium Noise Measurement with an On-chip Resonant Circuit}
\label{NoiseJoseph}
The design of the experiment allows to couple a noise source to the detector via the resonant circuit. For the sample discussed in this work, we couple a Josephson junction (see section \ref{SFAMT}) and measure its noise power of quasiparticles tunneling in the quantum regime ($h\nu>>k_BT$).
The interest in coupling a Josephson junction is twofold. First, it presents a strongly nonlinear I(V) characteristic responsible for a strong non linearity in the noise spectrum together with a strong difference between emission and absorption processes. Second, one can use its ac Josephson effect for calibration. In the following, we first calibrate the source/detector coupling and then measure the quasiparticles emission shot noise of the source Josephson junction.

\subsection{Calibration of the source/detector coupling $|Z_t|^2$ \textit{i.e.} the transimpedance}
\label{CalibZt}

\begin{figure}[htbp]
	\begin{center}
		\includegraphics[width=8cm]{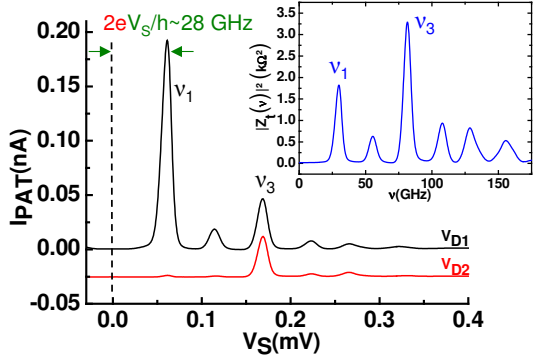}
	\end{center}
	\caption{For calibration, we measure at different bias voltage $V_D$ the PAT current through the detector versus the source junction bias $V_S$, when the source is in the regime of ac Josephson effect. The curves are shifted vertically for clarity. $V_D$ selects the noise frequencies $\nu$ of interest : the upper curve is taken at $V_{D1}=450 \mu$V, corresponding to $\nu \geq (2\Delta-eV_{D1})/h=17GHz$ whereas the lower curve, at $V_{D2}=300 \mu$V, corresponds to $\nu \geq 53$GHz. Inset : Frequency dependence of the coupling $|Z_t(\nu)|^2$ deduced from the curve taken at $V_{D1}$.}
	\label{fig3}
\end{figure}

When the source junction is biased it emits noise, which couples to the detector \textit{via} the resonant circuit. Consequently the coupling is efficient only close to the resonant frequencies of the resonator. In this part, we extract the coupling between source and detector \textit{i.e.} the transimpedance $Z_t(\nu)$. If the detector is voltage biased below the gap, it is essentially sensitive to voltage emission noise $S_V(-\nu)$. In this case, part of the current fluctuations of the source junction go through the coupling circuit. This leads to voltage fluctuations across the detector proportional to the transimpedance $Z_t(\nu)$. This quantity is the ratio between voltage fluctuations across the detector and current fluctuations emitted by the source $S_V(\nu)=|Z_t(\nu)|^2S_I(\nu)$. To probe $|Z_t(\nu)|^2$ the detector is voltage biased in the subgap region to be sensitive only to emission and we use the ac Josephson effect of the source junction for calibration (see Table \ref{SummaryConditions}). On figure \ref{fig3} the PAT current through the detector versus the source bias voltage $V_S$ is shown at two detector voltages $V_{D1}= 450 \mu$V and $V_{D2}= 300 \mu$V. Those positions allow to select frequencies of interest as will see in subsection \ref{SNM}. In this regime where the detector is irradiated by the Josephson effect at frequency $\nu=2eV_S/h$ the PAT current theoretically reads \cite{tien63}~: 
\begin{equation}
I_{PAT}(V_D)=\frac{1}{4}\frac{e^2|Z_t(\nu)|^2 I_C^2}{(h\nu)^2}I_{QP}(V_D+h\nu/e) 
\label{IjAC}
\end{equation}
with $I_C$ the critical current of the source junction and $I_{QP}(V_D)$ the $IV$ characteristic of the detector. Using this formula we can extract from the PAT current measured at $V_{D1}$ the value of the coupling $|Z_t(\nu)|^2$ (inset of fig. \ref{fig3}). It exhibits resonances at the same frequencies as the resonator. This detection scheme is characterized by a strong coupling proportional to the quality factor of the resonances of $|Z_t(\nu)|^2$, for a finite value of frequencies. This contrasts with previous experiments using a capacitive coupling between source and detector \cite{deblock03,billangeon06} which leads to a relatively small coupling over a wide range of frequencies.

In previous section \ref{SFAMT} we have emphasized the difference between the impedance seen by the detector $Z$ and the intrinsic impedance of the resonator $Z_r$ in the matrix impedance. We now compare $Z_{t,m}$, the transimpedance experimentally measured and $Z_t$ the intrinsic transimpedance in the matrix impedance. With the same type of calculations one has:
\begin{equation}
		Z_{t,m}=\frac{Z_t}{1+Z_r.[Y_S+Y_D]+Y_S.Y_D.[Z_r^2-Z_t^2]}.
	\label{eqZtmeasVsZtMatrix}
\end{equation}

Equation \ref{eqZtmeasVsZtMatrix} describes the evolution of the transimpedance as a function of $Y_S$, $Y_D$ and $Z_r$. In the case where admittances $Y_S$ and $Y_D$ are small ($Y_S,Y_D<<1/Z_r,1/Z_t$), we find that the measured transimpedance is exactly the transimpedance of the impedance matrix. Indeed, in this case the resonator is only slightly perturbed so that no difference can be seen between these two quantities. In the other limit where $Y_S$ and $Y_D$ are large compared to $1/Z_r$ and $1/Z_t$, the effective transimpedance vanishes because the resonator is short circuited by the source impedance or the detector. We finally conclude that differences between the amplitude of the effective transimpedance measured and the transimpedance of the resonator alone are small in the case of relatively high impedance nanodevices. However, as already mentioned, junctions also have capacitances which shift resonances to lowest frequencies. This resonance shift constitutes the main difference between $Z_{t,m}$  and $Z_{t}$.

\subsection{Out-of-equilibrium emission shot noise measurement}
\label{SNM}
We have seen in section \ref{CalibZt} that we are able to calibrate the coupling between the source and the detector. In this section we use it to quantitatively measure the emission noise associated with the tunneling of quasiparticles when the source junction is biased on the quasiparticles branch ($e V_S \geq 2 \Delta$). To do so, we apply a dc voltage bias to the detector in the emission noise sensitive region ($e V_D \leq 2 \Delta$) while sweeping the dc voltage source close to its quasiparticle branch (see Table \ref{SummaryConditions}). In addition of this dc is applied a small ac voltage ($\approx 3\mu V$ at $13.33$ Hz) and the modulated current through the detector is measured with a lock-in amplifier technique. The measured quantity is the derivative of the photoassisted current with respect to the source bias $\partial I_{PAT}/ \partial V_S$. Such a detection scheme is not sensitive to other noise contributions beside the source and increases the sensitivity of the measurement. 

\begin{figure}[htbp]
	\begin{center}
		\includegraphics[width=8cm]{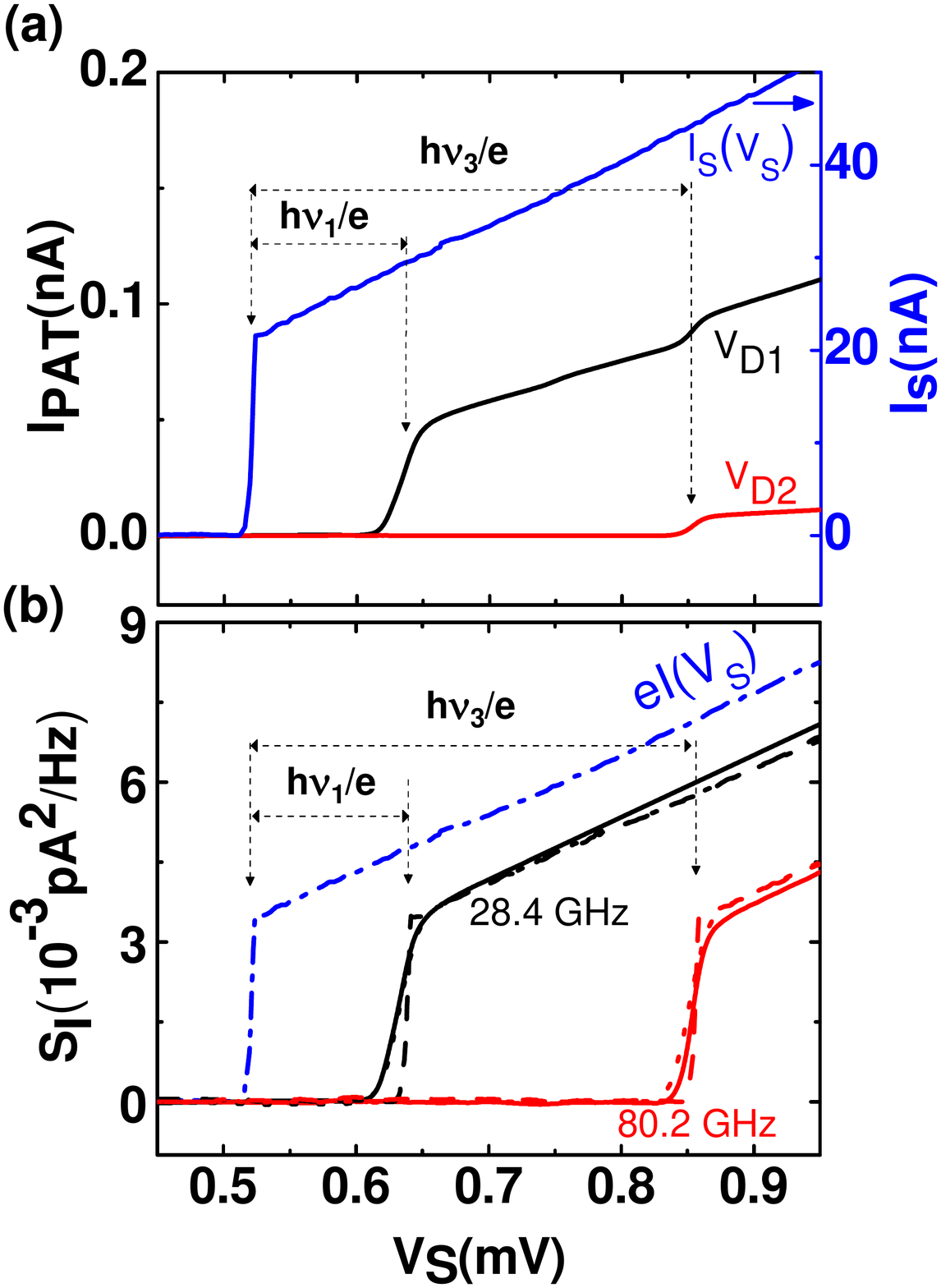}
	\end{center}
	\caption{(a) PAT current through the detector versus the source junction bias $V_S$, when the source is biased on the quasiparticle branch. The two curves are taken at $V_{D1}$ and $V_{D2}$. The $I(V)$ characteristic of the source junction is superimposed on the graph. (b) Extracted noise power in emission at $\nu_1=28.4$GHz and at $\nu_3=80.2$GHz. For comparison the expected noise power is plotted (dashed curve) together with the noise average over the bandwidth $\delta\nu_n$ of detection (dotted line). The agreement of this latter quantity is within 5\% with the extracted noise power. The expected zero frequency emission noise power is plotted in dashed-dotted line.}
	\label{fig4}
\end{figure}
The PAT current is obtained by a numerical integration of the signal. It is shown on figure \ref{fig4} for two values of detector bias $V_{D1}$ and $V_{D2}$. We use eq. \ref{IpatEm}, with $S_V(\nu)=|Z_t(\nu)|^2 S_{I_{QP}}(\nu,V_S)$ and $\delta \nu_n$ the width of the resonances of $|Z_t(\nu)|^2$ in order to extract quantitatively the noise spectrum from data in fig.\ref{fig4}. When the detector is biased at $V_D < 2 \Delta /e$, only the frequencies higher than $(2 \Delta -e V_D)/h$ have to be considered. Consequently for $V_D=V_{D1}$ the detector is mainly sensitive to the noise at frequencies $\nu_1$ and $\nu_3$, whereas for $V_D=V_{D2}$ only the noise at frequency $\nu_3$ is detected. The noise at $\nu_1$ is extracted from the curve $I_{PAT}(V_{D1})-\alpha I_{PAT}(V_{D2})$ with $\alpha$ ($>1$) a constant taking into account differences in sensitivity at frequency $\nu_3$ for the two detector positions. $\alpha$ is the ratio of the area of the peaks around $\nu=\nu_3$ at $V_D=V_{D2}$ and $V_D=V_{D1}$ (see fig.\ref{fig3}). One finally obtains the spectral density of quasiparticles noise in emission at $\nu_1=28.4$GHz and $\nu_3=80.2$GHz (Fig. \ref{fig4}). 
We compare these results to the theoretical prediction \cite{dahm69,rogovin74,billangeon06}:
\begin{equation}
S_{I_{QP}}(\nu,V_S)=e\left[\frac{I_{QP}(h\nu/e+V_S)}{1-e^{-\beta(h\nu+eV_S)}}+\frac{I_{QP}(h\nu/e-V_S)}{1-e^{-\beta(h\nu-eV_S)}}\right] 
\label{SIsis0}
\end{equation}

with $\beta=1/k_BT$ and to the noise integrated over the detection bandwidth $\delta \nu_n$. The agreement is within $5\%$ in amplitude with this last quantity and the frequency dependence is well reproduced. To our knowledge this is the first direct quantitative measurement in the quantum regime $h\nu>>k_BT$ of the noise associated with the quasiparticles tunneling. 

The sensitivity of our detection scheme is essentially limited by noise of the room temperature amplifiers which leads to a lowest measurable current of $20fA/\sqrt{Hz}$. This gives a minimum measurable current noise, with this setup, of $2fA^2/Hz$ at $28GHz$ and $8fA^2/Hz$ at $80GHz$. If we convert this result in terms of noise temperature $T_N$ accross a $20 k\Omega$ resistor, we get respectively $T_N(28GHz)=1.5mK$ and $T_N(80GHz)= 5.8 mK$. We stress here that this detection scheme only works for high impedance nanodevices.

\subsection{Out-of-equilibrium absorption noise measurement}

The detection principle of the absorption noise is the same as in the previous section with the detector junction biased on the quasiparticles branch. However, in this configuration, the direct extraction of the noise spectra is very delicate due to the simultaneous contribution of the absorption noise of the source and the resonator which is itself influenced by the impedance of the source \cite{billangeon06}. As a result, the measured signal consists in a non trivial combination of the absorption noise of the resonator and the source which prevents for a reasonable extraction of the out-of-equilibrium absorption noise and as a consequence the verification of the generalized fluctuation-dissipation theorem.

\section{Equilibrium Noise Measurement with an On-chip Resonant Circuit}
\label{QNMWOCRC2}

In this section we show how using the setup described previously, one can measure the high frequency quantum noise of the resonator at \textit{equilibrium}. This is done by measuring the dc $I(V)$ characteristic of the detector which is modified by the presence of the noisy resonator.

\subsection{Theoretical expectation for the noise of the resonator}
\label{NoiseRes}
According to the Quantum Fluctuation-Dissipation Theorem (QFDT), the voltage fluctuations of a resistive circuit at equilibrium of impedance $Z(\nu)$ reads \cite{kubo66,clerk10}: 
\begin{equation}
S_V(\nu,T)=\frac{2Re[Z(\nu)]h\nu}{1-\exp{(-h\nu/k_B T)}}
\label{SV}	
\end{equation}
 This formula describes the crossover between thermal noise at low frequency and quantum noise related to the zero point fluctuations of the electromagnetic field at frequency higher than $k_BT$.

\begin{figure}[htbp]
	\begin{center}
		\includegraphics[width=8cm]{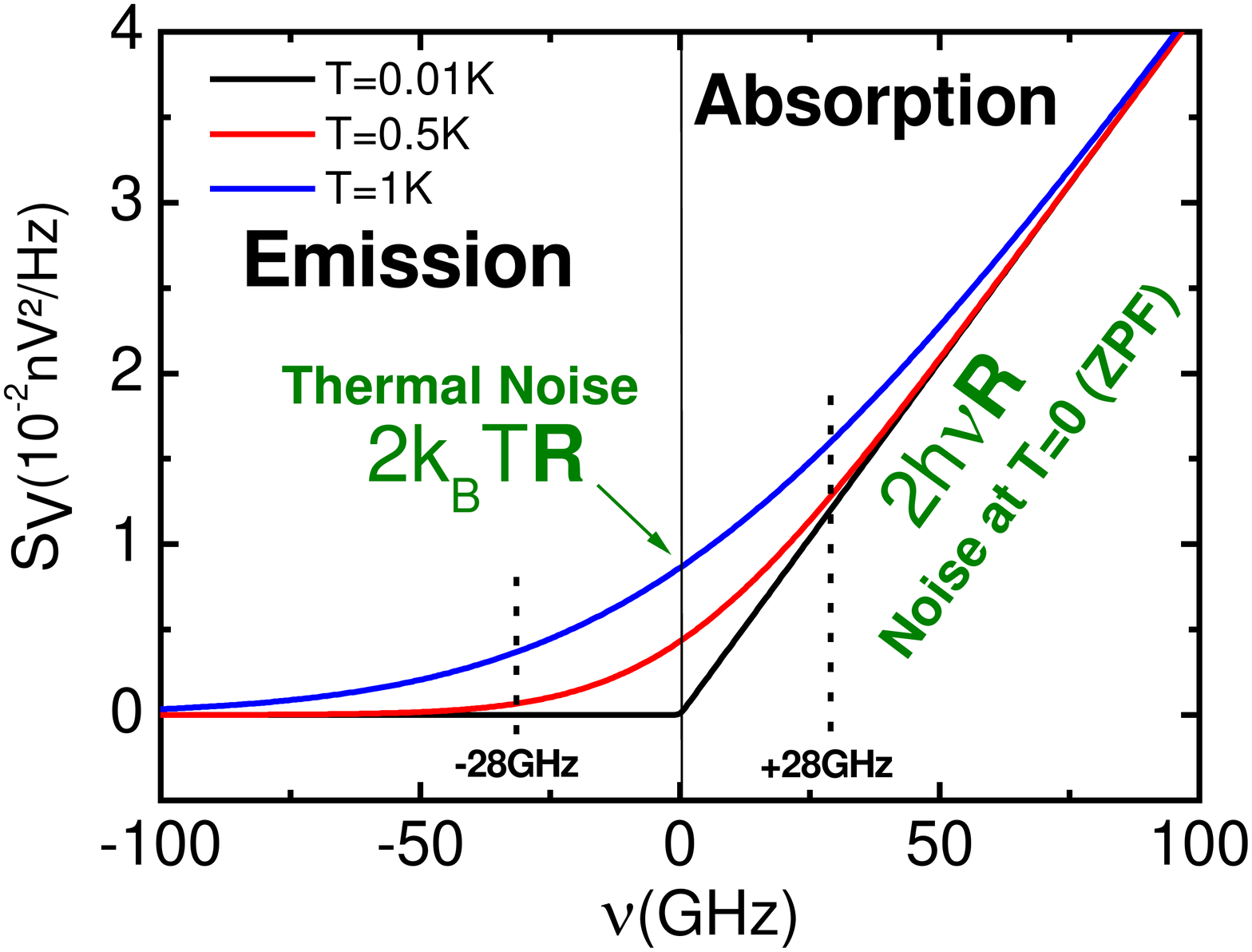}
	\end{center}
	\caption{Expected frequency dependence of the equilibrium voltage noise of a resistor $R=300\Omega$ at three different temperatures (formula \ref{SV}). At $\nu=0$, the prediction follows the Johnson-Nyquist formula for a non symmetrized thermal noise. At $T=0$, there is no emission noise whereas the absorption noise increases linearly with the frequency. This noise at $T=0$ is related to the zero point fluctuations of the voltage bias accross the resistor at equilibrium}
	\label{SvThnuResonator}
\end{figure}
In the specific case of the resonator, the resonances seen in $Re[Z(\nu)]$ at $\nu_n$ give rise to noise peaks at frequencies $+\nu_n$ (absorption, with $\nu_n >0$) and $-\nu_n$ (emission). At low temperature only peaks in absorption are predicted whereas when the temperature increases peaks in emission should appear (see Fig.\ref{SvThnuResonator}).

\subsection{Measurement of $dI_D/dV_D$ as a probe of voltage noise $S_V$}

We have seen in sections \ref{SFAMT} and \ref{QA} that the subgap $I(V)$ characteristic of the detector related to tunneling of Cooper pairs (low bias in the ac Josephson effect range) is modified by the resonant environment. Hereafter we demonstrate that the resonant circuit coupled to the detector junction also has an effect on the quasiparticles branch. We measure the $dI/dV$ rather than $I(V)$ characteristic of the detector junction (Fig. \ref{fig2}a) to increase the sensitivity of the experiment (see Table \ref{SummaryConditions}). At low temperature, on top of the expected $dI/dV$ curve of the detector, we see peaks (denoted by arrows on figure \ref{fig2}) at bias voltages $eV_D=2 \Delta + h\nu_n$ with $\nu_n$ the resonant frequency of the circuit coupled to the detector. These peaks are not detected for bias voltage below $2 \Delta$. This is not true at higher temperature where a peak at $e V_D=2 \Delta -h \nu_1$ appears and grows with temperature between 20mK and 1K. The position in $V_D$ of the peaks changes due to the temperature dependence of the superconducting gap. Higher temperature were not considered due to the strong temperature dependence of the SIS detector for $T > 1$K. These peaks in the $dI/dV$ characteristics of the detector are attributed to its sensitivity to the voltage fluctuations of the resonant circuit. Hereafter, we treat the data to extract emission and absorption noise power at the resonant frequencies of the on-chip circuit.  
 
\subsection{Extraction of the equilibrium noise of the resonator}

The noise extraction is based on Eq.\ref{IPAT}. For $V_D < 2\Delta /e$ only the first term in Eq.\ref{IPAT} is non-zero. For an emission noise peaked at frequencies $-\nu_n$, approximating the integral by a sum yields~:
\begin{equation}
	I_{PAT}(V_D)=\sum_n \left( \frac{e}{h\nu_n} \right)^2 S_V(-\nu_n) \delta\nu_n I_{QP,0}(V_D+\frac{h\nu_n}{e})
	\label{IpatEm}
\end{equation}
with $\delta \nu_n=1.06 \nu_n/Q_n$ related to the width of the resonance at frequency $\nu_n$, which can be extracted from Fig. \ref{ReZseul} ($\delta \nu_n$ multiplied by the amplitude of the resonance peak is equal to the area of the peak). On the other hand for $V_D > 2 \Delta/e$, only the absorption term in Eq.\ref{IPAT} leads to peaks in $dI/dV$ at $V_D=(2 \Delta + h \nu_n)/e$~: 
\begin{equation}
	I_{PAT}(V_D)=\sum_n \left( \frac{e}{h\nu_n} \right)^2 S_V(\nu_n) \delta\nu_n I_{QP,0}(V_D-\frac{h\nu_n}{e})
	\label{IpatAb}
\end{equation}

\begin{figure}[htbp]
	\begin{center}
		\includegraphics[width=8cm]{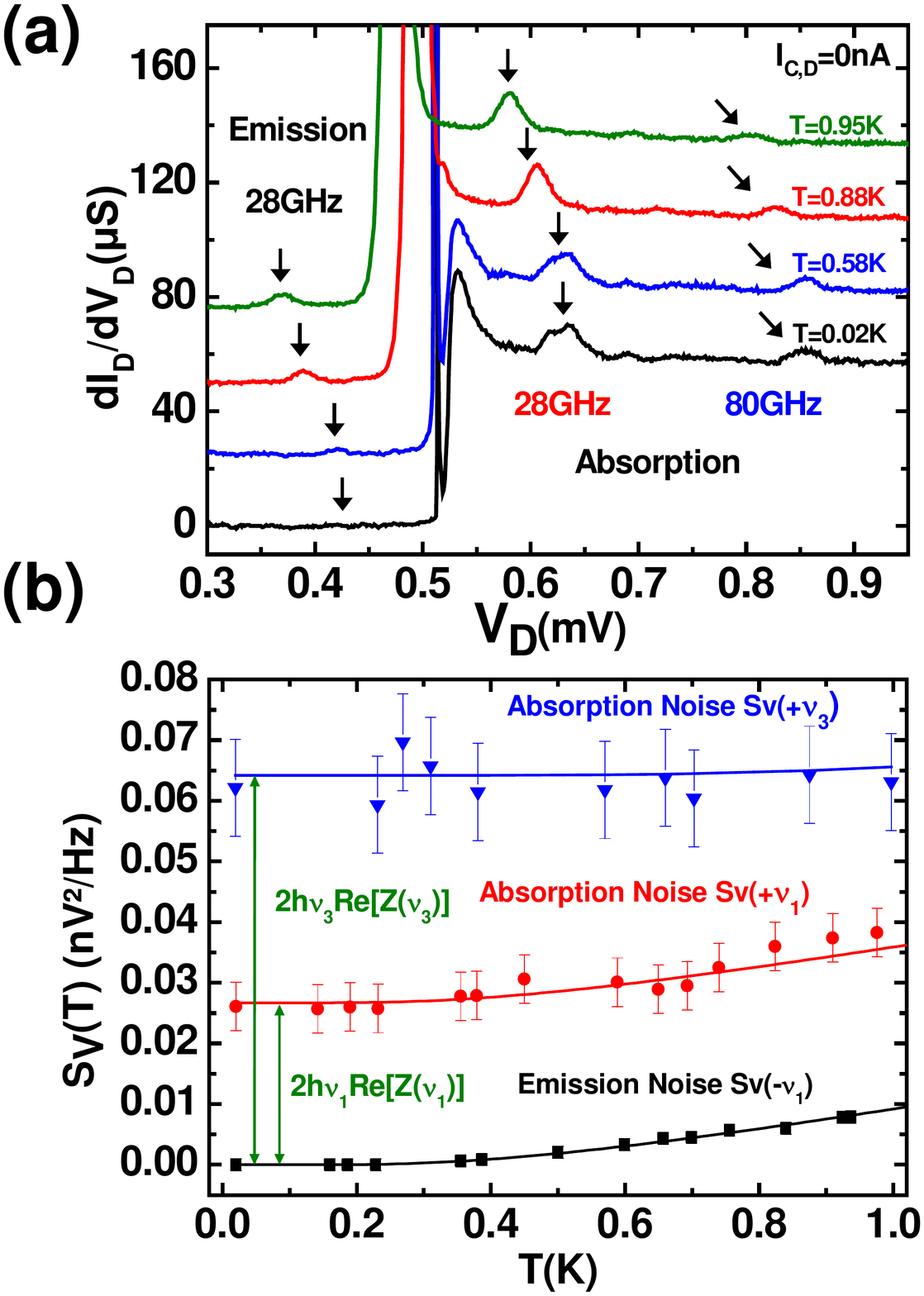}
	\end{center}
	\caption{ (a) Differential conductance $dI/dV_D$ of the detector junction at different temperatures with $I_C$ minimized by adjusting the magnetic flux. The curves are shifted vertically for clarity. The peaks corresponding to the detection of emission or absorption noise are denoted by arrows. (b) Dependence versus temperature of the power of voltage noise at $\nu_1=28.4GHz$ and $\nu_3=80GHz$ in emission and in absorption. The solid lines correspond to the theoretical predictions (Eq.\ref{SV}) with $Re[Z(\nu_1)]=714\Omega$ and $Re[Z(\nu_3)]=604\Omega$. Only absorption noise, consistent with zero point fluctuations, is detected below 0.4 K.}
	\label{fig2}
\end{figure}

From these relations we extract quantitatively the emission and absorption voltage fluctuations of the resonant circuit at $\nu_1=28.4$ GHz for $T$ between 20mK and 1K (Fig.\ref{fig2}b). To do so we integrate the corresponding peak in $dI/dV_D$, at $V_D=(2 \Delta - h \nu_1)/e$ for emission and $V_D=(2 \Delta + h \nu_1)/e$ for absorption, to obtain the value of $I_{PAT}$. An important point concerning the absorption noise extraction is to substract carefully the baseline due to elastic quasiparticles tunneling in the $dI/dV_D$ curves. This is the main reason why error bars are larger for absorption than for emission. $\delta \nu_1=6.66$ GHz is extracted from fig.\ref{ReZseul} and the current $I_{QP,0}(V)$ is measured from the $I(V)$ characteristic of the detector at temperature $T$. The same treatment can be done for the absorption noise at $\nu_3=80.2$Ghz and leads to $S_V(\nu_3)=0.062 \pm 0.008$nV$^2/$Hz between 20 and 700 mK, consistent with the expected value of $0.064$nV$^2/$Hz. The emission noise has not been measured at $\nu_3$ because the expected signal is too small at these temperatures compared to the noise of the experimental setup. In addition, to look at the expected peaks it is necessary to bias the detector at smaller voltages where the ac Josephson peaks are not completely suppressed by magnetic flux. The temperature dependence of voltage fluctuations agrees quantitatively with theoretical predictions (Eq.\ref{SV}) using the calibration of $Re[Z(\nu)]$ described in section \ref{SFAMT}. Indeed, deep in the quantum regime, when $h \nu_1 \gg k_B T$, the voltage fluctuations at \textit{equilibrium} of the circuit do not exhibit any emission noise whereas as a result of the zero point fluctuations of the electromagnetic field the circuit still shows absorption noise. In the intermediate regime, when $h\nu_1 \geq k_B T$ we see the crossover to thermal noise.

\section{Conclusion}
In this paper we have shown that by coupling a quantum detector, a SIS junction, to a noise source through a resonant circuit it is possible to measure the equilibrium noise of the resonant coupling circuit, the out-of-equilibrium noise of a Josephson junction and the finite frequency admittance of the same Josephson junction. The equilibrium noise of the resonator exhibits a strong asymmetry between emission and absorption related to zero point fluctuations. The out-of-equilibrium emission noise of quasiparticles tunneling of the Josephson junction shows a strong frequency dependence in agreement with theoretical predictions and finally the quantum admittance components highlight frequency dependent singularities. Finally, the technique described above can be used to probe the dynamics of other relatively resistive correlated mesoscopic systems at high frequency. This was recently achieved for the noise measurement of a carbon nanotube quantum dot in the Kondo regime \cite{bassetKondo11}.

\begin{table*}[htbp]
	\begin{center}
		\begin{tabular}{|p{2cm}||p{2cm}|p{2cm}|p{3cm}|p{3cm}|p{2cm}|}
		\hline
		\textbf{Extracted quantity} & \textbf{Critical current of the detector} $I_{C,D}$ & \textbf{Critical current of the source} $I_{C,S}$ & \textbf{Detector polarisation} & \textbf{Source polarisation} & \textbf{Measured quantities} \\
		\hline
		\hline
		\textbf{$Re[Z(\nu)]$} & Maximum & Minimum & sweep $V_D\in[-2\Delta/e,2\Delta/e]$ & fix $V_S=0$ & $dI_D/dV_D$, $I_D(V_D)$\\
		\hline
		\textbf{$Z_t(\nu)$} & Minimum & Maximum & fix $V_D\in\pm[(2\Delta-h\nu)/e,2\Delta/e]$ & sweep $V_S\in[-2\Delta/e,2\Delta/e]$ & $dI_D/dV_S$, $I_D(V_S)$\\
		\hline
		\textbf{$S_V$ resonator} & Minimum & Minimum & sweep $V_D\in\pm[(2\Delta-h\nu)/e,(2\Delta+h\nu)/e]$ & fix $V_S=0$ & $dI_D/dV_D$, $I_D(V_D)$\\
		\hline
		\textbf{$S_I$ source} & Minimum & Minimum & fix $V_D\in\pm[(2\Delta-h\nu)/e,2\Delta/e]$ & sweep $V_S\in\pm[(2\Delta-h\nu)/e,(2\Delta+h\nu)/e]$ & $dI_D/dV_S$, $I_D(V_S)$\\
		\hline
		\textbf{Quantum Conductance} $ReY_S$ & Maximum & Minimum & fix $V_D$ bottom resonant peak & sweep $V_S\in\pm[(2\Delta-h\nu)/e,(2\Delta+h\nu)/e]$ & $dI_D/dV_S$, $I_D(V_S)$\\
		\hline
		\textbf{Quantum Susceptance} $ImY_S$ & Maximum & Minimum & fix $I_D$ inflection point of resonant peak & sweep $V_S\in\pm[(2\Delta-h\nu)/e,(2\Delta+h\nu)/e]$ & $dV_D/dV_S$, $V_D(V_S)$\\
		\hline
		\end{tabular}
	\end{center}
\caption{Summary of the different experimental conditions used to measure interesting quantities.}
\label{SummaryConditions}
\end{table*}

\section{Acknowledgements}
We acknowledge M. Aprili, B. Reulet, J. Gabelli, I. Safi, P. Simon, F. Portier, M. Hoffheinz, C. Ojeda-Aristizabal, B. Dassonneville, M. Ferrier and S. Gu{\'e}ron for fruitful discussions. This work has benefited from financial support of ANR under project DOCFLUC (ANR-09-BLAN-0199-01) and C'Nano Ile de France under project HYNANO.

\end{document}